\documentclass[12pt]{article}

\usepackage{amsmath}
\usepackage{amssymb}
\usepackage{epsfig}
\usepackage{graphicx}
\usepackage{cite}
\usepackage{multirow}
\usepackage{longtable}
\usepackage{lscape}
\usepackage{bbm}
\usepackage{color}

\newcommand{\capdef}{}
\newcommand{\mycaption}[2][\capdef]{\renewcommand{\capdef}{#2}%
        \caption[#1]{{\footnotesize #2}}}
\makeatletter
\renewcommand{\fnum@table}{\textbf{\tablename~\thetable}}
\renewcommand{\fnum@figure}{\textbf{\figurename~\thefigure}}
\makeatother

\newcounter{myenumi}

\renewcommand{\themyenumi}{\roman{myenumi}}
{\end{list}}

\newlength{\myem}
\newcounter{mysubequation}[equation]

\makeatletter
\renewcommand{\section}{\@startsection{section}{1}{0em}{-\baselineskip}%
{\baselineskip}{\normalfont\large\bfseries}}
\renewcommand{\subsection}%
{\@startsection{subsection}{2}{0em}{-0.7\baselineskip}%
{0.7\baselineskip}{\normalfont\bfseries}}
\makeatother

\newcommand{\bi}{\begin{itemize}}
\newcommand{\ei}{\end{itemize}}

\newcommand{\be}{\begin{equation}}
\newcommand{\ee}{\end{equation}}
\newcommand{\bea}{\begin{eqnarray}}
\newcommand{\eea}{\end{eqnarray}}

\newcommand{\deltacp}{\delta_{\mathrm{CP}}}
\newcommand{\stheta}{\sin^2 2 \theta_{13}}

\newcommand{\ie}{{\it i.e.}}

\newcommand{\eg}{{\it e.g.}}

\newcommand{\cf}{{\it cf.}}

\newcommand{\eq}{Eq.}
\newcommand{\eqs}{Eqs.}

\newcommand{\fig}{Fig.}

\newcommand{\Ref}{Ref.}
\newcommand{\Refs}{Refs.}
\newcommand{\Sec}{Sec.}
\newcommand{\Secs}{Secs.}
\newcommand{\App}{App.}

\newcommand{\Tab}{Table}

\newcommand{\equ}[1]{\eq~(\ref{equ:#1})}
\newcommand{\equt}[2]{\eqs~(\ref{equ:#1}) and~(\ref{equ:#2})}

\newcommand{\figu}[1]{\fig~\ref{fig:#1}}

\begin{document}

\begin{titlepage}

\renewcommand{\thefootnote}{\alph{footnote}}

\vspace*{-3.cm}
\begin{flushright}
\end{flushright}


\renewcommand{\thefootnote}{\fnsymbol{footnote}}
\setcounter{footnote}{-1}

{\begin{center}
{\large\bf
Entangled maximal mixings in $\boldsymbol{U_{\mathrm{PMNS}}=U_\ell^\dagger U_\nu}$, \\
and a connection to complex mass textures} \end{center}}
\renewcommand{\thefootnote}{\alph{footnote}}

\vspace*{.8cm}
\vspace*{.3cm}
{\begin{center} {\large{\sc
 		Svenja~Niehage\footnote[1]{\makebox[1.cm]{Email:}
                svenja.niehage@physik.uni-wuerzburg.de},
                Walter~Winter\footnote[2]{\makebox[1.cm]{Email:}
                winter@physik.uni-wuerzburg.de}
                }}
\end{center}}
\vspace*{0cm}
{\it
\begin{center}

       Institut f{\"u}r Theoretische Physik und Astrophysik, Universit{\"a}t W{\"u}rzburg, \\
       D-97074 W{\"u}rzburg, Germany

\end{center}}

\vspace*{1.5cm}

{\Large \bf
\begin{center} Abstract \end{center}  }

We discuss two different configurations of $U_{\mathrm{PMNS}}=U_\ell^\dagger U_\nu$
with maximal mixings in {\em both} $U_\ell$ and $U_\nu$. The non-maximal mixing angles are assumed to be small, which means that they can be expanded in. Since we are particularly interested in the implications for CP violation, we fully take into account complex phases.
We demonstrate that one possibility leads to intrinsically large $\stheta$ and strong deviations from maximal mixings. The other possibility is generically close to tri-bimaximal mixing, and allows for large CP violation.  We demonstrate how the determination of the $\theta_{23}$ octant and the precision measurement of $\deltacp$ could discriminate among different qualitative cases.
In order to constrain the unphysical and observable phases even further, we relate our configurations to complex mass matrix textures.  In particular, we focus on phase patterns which could be generated by powers of a single complex quantity $\eta \simeq \theta_C \, \exp(i \, \Phi)$, which can be motivated by Froggatt-Nielsen-like models.
For example, it turns out that in all of the discussed cases, one of the Majorana phases is proportional to $\Phi$ to leading order.
In the entire study, we encounter three different classes of sum rules, which we systematically classify.

\vspace*{.5cm}

\end{titlepage}

\newpage

\renewcommand{\thefootnote}{\arabic{footnote}}
\setcounter{footnote}{0}

\section{Introduction}

For Majorana neutrinos and the effective $3 \times 3$-case,
the neutrino mass matrix can be diagonalized by (here, we closely follow \Refs~\cite{Frampton:2004ud,Plentinger:2006nb,Plentinger:2007px}) 
\begin{equation}\label{equ:Maj}
 M_\nu^\text{Maj}=U_\nu M_\nu^\text{diag}U_\nu^T \, ,
\end{equation}
and the charged lepton mass matrix by
\begin{equation}
 \label{equ:lepdiag}
M_\ell=U_\ell\,M_\ell^\text{diag}\,{U_\ell'}^\dagger,
\end{equation}
where $U_\nu$, $U_\ell$, and $U_\ell'$ are general unitary matrices.
The (real) eigenvalues can be written as
$M_\ell^\text{diag} = \text{diag}(m_e,m_\mu,m_\tau)$
and
$M_\nu^\text{diag}=\text{diag}(m_1,m_2,m_3)$,
and are, except from the absolute neutrino mass scale,
experimentally known. 
The neutrino mixing matrix $U_{\mathrm{PMNS}}$ is generated as a product of two matrices
$U_\ell$ and $U_\nu$ which enter the charged current interaction Lagrangian, \ie,
$U_{\mathrm{PMNS}}=U_\ell^\dagger U_\nu$. 
Therefore, $U_{\mathrm{PMNS}}$ describes the relative rotation
between the charged lepton $U_\ell$ and the neutrino mixing $U_\nu$ matrices, which turn the left-handed fields into the respective mass bases. Note that the mass matrix entries usually emerge as a product of the relevant Yukawa couplings and the vacuum expectation value of the Higgs. In addition, note that 
once the neutrino masses are known, the Majorana mass matrix $M_\nu^\text{Maj}$ can be reconstructed from 
$U_\nu$ according to \equ{Maj}. However, for the charged lepton mass matrix,
an additional mixing matrix $U_\ell'$ turning the right-handed fields appears. Therefore,
there is much more freedom involved in the reconstruction of $M_\ell$, and we hence
do not discuss specific structures for $M_\ell$.

While the physical observables in $U_{\mathrm{PMNS}}$ are independent of the basis
used for \equt{Maj}{lepdiag}, theoretical models often use the structure of the mass matrices
as a starting point. These structures are often called ``textures''. They are obviously basis dependent,
and often predicted within a more general theoretical framework.
For example, masses for quarks and leptons may arise from higher-dimension terms via
the Froggatt-Nielsen mechanism~\cite{Froggatt:1978nt} in combination with a flavor symmetry:
\begin{equation}
\mathcal{L}_\mathrm{eff} \sim K_{ij} \, \langle H \rangle \, \eta^{n_{ij}} \, \bar{\Psi}_L^i \Psi_R^j \,  + h.c. \, .
\label{equ:fn}
\end{equation}
In this case, $\eta$ is a small
parameter $\eta=v/M_F$ which controls the flavor symmetry breaking. 
Here $v$ are universal VEVs of SM (Standard Model) singlet scalar ``flavons'' that break the flavor symmetry,
and $M_F$ refers to the mass of super-heavy fermions, which are charged under the
flavor symmetry. The SM fermions are given by the $\Psi$'s, and $K_{ij}$ are order unity (complex) numbers.
The integer powers $n_{ij}$ are solely determined by the
quantum numbers of the fermions under the flavor symmetry (see, \eg, \Refs~\cite{Enkhbat:2005xb,Plentinger:2007px}), and  
the order one coefficients $K_{ij}$ can be used to fit a particular texture to data. Obviously, the neutrino and charged lepton mass matrices in \equ{Maj} and \equ{lepdiag} are then, to a first approximation, described by the powers $M_{ij} \propto \eta^n_{ij}$, and $U_\ell$ or $U_\nu$ will, in general,  {\em not} be diagonal.
Therefore, large mixings may come from either the charged lepton or neutrino sector, as it has, for instance, been discussed in \Refs~\cite{Altarelli:2004jb,Romanino:2004ww,Antusch:2004re,deS.Pires:2004bq,Li:2005yj,Ohlsson:2005js,Hochmuth:2007wq}. In this study, however, we discuss large mixing angles in {\em both} $U_\ell$ and $U_\nu$, which lead to ``entangled'' maximal mixings (see also \Ref~\cite{Masina:2005hf}). In addition, it is obvious from  \equt{Maj}{lepdiag} that if $M_\nu^\text{Maj}$ and $M_\ell$
are both real, then $U_\mathrm{PMNS}$ will be real and there will be no leptonic CP violation. Therefore, we study the full complex case.

{\bf Entangled maximal mixings.}
In the first part of the study (\cf, \Sec~\ref{sec:obs}), we use the standard parameterization for a CKM-like matrix
\begin{equation}
 \label{equ:ckm}
 \widehat{U} = \left( 
 \begin{array}{ccc}
   c_{12} c_{13} & s_{12} c_{13} & s_{13} e^{-\text{i}\widehat{\delta}} \\
   -s_{12} c_{23} - c_{12} s_{23} s_{13} e^{\text{i}\widehat{\delta}} &   c_{12} c_{23} -
 s_{12} s_{23} s_{13} e^{\text{i}\widehat{\delta}} & s_{23} c_{13} \\ 
 s_{12} s_{23} - c_{12} c_{23} s_{13}
e^{\text{i}\widehat{\delta}} & -c_{12} s_{23} - s_{12} c_{23} s_{13} e^{\text{i}\widehat{\delta}} & c_{23}
c_{13} 
 \end{array}
 \right) 
\end{equation}
for both $U_\ell$ and $U_\nu$,
where $s_{ij} = \sin \widehat{\theta}_{ij}$ and 
$c_{ij} = \cos \widehat{\theta}_{ij}$. We adopt the purely phenomenological point of view that a number of mixing angles in $U_\ell$ and $U_\nu$  are maximal, and the others are small. The maximal mixing angles are assumed to be exactly maximal because they might be generated by a symmetry,
and the small mixing angles are assumed to be $\mathcal{O}(\theta_C)$, which allows us to expand in them.
We call a particular combination of maximal and small mixing angles a ``configuration''. We study two of these configurations in \Sec~\ref{sec:obs}, where the stars refer to the small mixing angles:
\begin{description}
\item[Configuration~1] with three max. mixing angles $(\theta_{12}^\ell, \theta_{13}^\ell, \theta_{23}^\ell, \theta_{12}^\nu, \theta_{13}^\nu, \theta_{23}^\nu) = (*,\frac{\pi}{4}, \frac{\pi}{4}, *, \frac{\pi}{4}, *)$
\item[Configuration~2] with two max. mixing angles $(\theta_{12}^\ell, \theta_{13}^\ell, \theta_{23}^\ell, \theta_{12}^\nu, \theta_{13}^\nu, \theta_{23}^\nu)=(*,*, \frac{\pi}{4}, \frac{\pi}{4}, *, *)$
\end{description}
As we will demonstrate, both of these configurations lead to valid implementations of $U_{\mathrm{PMNS}}$. We will use the currently allowed ranges for the observables to constrain the relationship between unphysical mixing angles and phases, and we will derive sum rules relating the observables. 

{\bf Connection to quark sector and complex mass matrices}.
In order to establish a connection between the quark and lepton sectors, quark-lepton complementarity (QLC)~\cite{Petcov:1993rk,Smirnov:2004ju,Raidal:2004iw,Minakata:2004xt} has been proposed as a
possibility to account for the differences between the quark and
lepton mixings. In QLC,
the quark and lepton mixing angles are connected by QLC relations such as
\begin{equation}\label{equ:qlc}
\theta_{12}+\theta_\text{C}\simeq\pi/4 \, .
\end{equation}
Sum rules such as \equ{qlc}
do not only include observables from the lepton sector, but also the Cabibbo angle.
They can be obtained if the mixing among the neutrinos and among the charged leptons is
described by maximal or CKM-like mixing angles, such as in $U_{\mathrm{PMNS}} \simeq V_\mathrm{CKM}^\dagger U_{\mathrm{bimax}}$ or similar constructions~\cite{Jezabek:1999ta,Giunti:2002pp,Frampton:2004ud,Antusch:2005kw,Ohlsson:2005js,Kang:2005as,Cheung:2005gq,Chauhan:2006im}. 
Instead of relying on special cases, extended quark-lepton complementarity (EQLC)~\cite{Plentinger:2006nb} is based on the hypothesis that all mixing angles can only be from the set $\{ \pi/4, \epsilon, \epsilon^2, \hdots \}$ with $\epsilon \simeq \theta_C$.\footnote{Other parameterizations of
$U_\text{PMNS}$ as a function of $\theta_\text{C}$ were, for instance, discussed in \Refs~\cite{Rodejohann:2003sc,Li:2005ir,Xing:2005ur,Datta:2005ci,Everett:2005ku}.}
 Quite naturally, 
entangled maximal mixing configurations emerge from this assumption. In addition, it is a nice feature of EQLC to predict the textures of the charged lepton and neutrino mass matrices to be of the form $M_{ij} \sim \epsilon^{n_{ij}}$ with $n_{ij} \in \{0,1,2, \hdots\}$, which allows for a direct connection to the Froggatt-Nielsen mechanism. Therefore, compared to many other approaches which focus on either the mixings or the hierarchies, EQLC aims to describe both the mixings and hierarchies.
EQLC has successfully been applied to the see-saw mechanism~\cite{Plentinger:2007px}, and the effective $3\times3$ case has been extended to complex mass matrices~\cite{Winter:2007yi}.  For the seesaw case and real $\eta=\epsilon \simeq \theta_C$,  a systematic connection to discrete flavor symmetries has been established in \Ref~\cite{Plentinger:2008up}, and a connection to SU(5) GUTs was studied in \Ref~\cite{Plentinger:2008nv}.
 The order of magnitude relations for the
neutrino masses for $M_\nu^\text{diag}$ in \equ{Maj} can in this context be written as
\begin{equation}
 m_1:m_2:m_3=\epsilon^2:\epsilon:1,\quad
 m_1:m_2:m_3=1:1:\epsilon,\quad m_1:m_2:m_3=1:1:1,
\label{equ:numassratios}
\end{equation}
where $m_1,m_2,$ and $m_3$, denote the masses of the 1st, 2nd, and 3rd neutrino
mass eigenstate. The 1st, 2nd, and 3rd equation in
\equ{numassratios} describe a
normal hierarchical (NH), inverse hierarchical (IH), and quasi
degenerate (QD) spectrum, respectively.\footnote{For NH neutrinos, one can compute 
$\epsilon$ from the current best-fit values, which gives $0.15 \lesssim \epsilon
\lesssim 0.22$ ($3\sigma$).} We choose the NH case as an example for this study.

In the second part of this study (\cf, \Sec~\ref{sec:matrix}), we extend the texture concept of \Refs~\cite{Plentinger:2006nb,Plentinger:2007px,Winter:2007yi} to the complex case. Similar to  \Ref~\cite{Plentinger:2006nb}, we 
parameterize the small mixing angles $\mathcal{O}(\epsilon)$  by powers of
$\epsilon \simeq \theta_C$, \ie,  $\epsilon$, $\epsilon^2$, and $0$. Here we assume that higher powers than two in the neutrino sector are absorbed in the current measurement precision, and we approximate these by zero.
We then define a texture by the leading orders in $\epsilon$ in the individual matrix entries, which are determined by the first non-vanishing coefficients.  Compared to earlier works, we do not only use the
absolute values of the leading order entries $\epsilon^n$, but also their phases $n \Phi$ (except from a global phase, which can be removed for each matrix). This means that the leading order entry
is given by $\eta^n=\epsilon^n \, \exp( i n \Phi)$.
For example, a texture may then read for the choice $\Phi=\pi/2$ as 
\begin{equation}
M_\nu^\mathrm{Maj} \sim
\left(
\begin{array}{ccc}
\eta & \eta & \eta^2 \\
\eta & \eta & 0 \\
\eta^2  & 0 & 1 \\
\end{array}
\right)
=
\left(
\begin{array}{ccc}
\epsilon \, e^{i \frac{\pi}{2}} & \epsilon \, e^{i \frac{\pi}{2}} & \epsilon^2 \, e^{i \pi} \\
\epsilon \, e^{i \frac{\pi}{2}} & \epsilon \, e^{i \frac{\pi}{2}} & 0 \\
\epsilon^2 \, e^{i \pi} & 0 & 1 \\
\end{array}
\right) \, ,
\label{equ:texture}
\end{equation}
where a ``0'' corresponds to $\mathcal{O}(\epsilon^3)$ with an undefined phase.
Note that in Froggatt-Nielsen-like (FN) models, one needs (at least) two standard model singlet flavon fields
with different $U(1)_{\mathrm{FN}}$ charges in order to produce CP violation, because for one fields the phase can be gauged away (see, \eg, \Refs~\cite{Lee:1974jb,Robinson:1997ub,Kanemura:2007yy}). In this case, for patterns, such as
in \equ{texture}, it is required that one field dominate for mild hierarchies, such as the neutrino mass hierarchy.
For example, one may have two different hierarchical VEVs, or two identical VEVs with very different $U(1)_{\mathrm{FN}}$ charges. In these cases, the neutrino mass matrix will be dominated by one field, but the other field can be present in the charged lepton and quark mass matrices (or in the texture zeros; \cf, \equ{texture}). The phase $\Phi$ will then be the relative phase between the two VEVs of the fields. In addition, the order one coefficients are assumed to be real (and positive), \cf, \equ{fn}.
The advantage of a texture definition such as \equ{texture}
is quite obvious: The condition to reproduce the correct powers
of the phases in the texture leads to implications for the unphysical phases and the observables. 
In \Sec~\ref{sec:matrix}, we study this condition in combination with the configurations from \Sec~\ref{sec:obs}, its effect on CP violation, and the emerging connection between the quark and lepton sectors via $\epsilon \simeq \theta_C$. 

Before we present our analysis in \Secs~\ref{sec:obs} and~\ref{sec:matrix}, we summarize in \Sec~\ref{sec:method} our notation and method. After our analysis, we summarize our results in \Sec~\ref{sec:summary}.

\section{Notation and method}
\label{sec:method}

In this section, we clarify our notation and describe some details of our method used in the following sections.

\subsection{Sum rules}
\label{sec:sumrules}

Throughout this study, we will encounter a number of different sum rules depending on the configuration studied and input used. In order to qualify as a sum rule, we require either a simple, testable connection among observables, which can be used to falsify our model (sum rule types~I and~II), or a simple relationship among observables and {\em one} unphysical quantity which allows for a straightforward extraction of this unphysical quantity (sum rule type~III). In addition, we require that there be only bare angles in our sum rules, \ie, that all trigonometric functions of the angles be expanded. For a systematic approach, we classify the sum rules according to the following scheme (with some examples from this study):
\begin{description}
\item[Type~I sum rules (lepton sector sum rules)]
 Sum rules relating lepton sector observables only; no unphysical quantities or model parameters are present in these sum rules. Example (\cf, \Sec~\ref{sec:t130}):
\begin{equation}
\theta_{23}+\frac{1}{2} \, \theta_{13}^2 \simeq \frac{\pi}{4} \, . \nonumber
\end{equation}
This type of sum rules can be used to falsify our model if {\em all} of the contributing
observables (at least two) are measured. It can be analogously translated in to the quark sector.
\item[Type~II sum rules (QLC-type sum rules)] 
 Sum rules relating lepton and quark sector observables; no unphysical quantities or model parameters are present in these sum rules (see, \eg, \Refs~\cite{Minakata:2004xt,King:2005bj,Plentinger:2008up}). Example (\cf, \Tab~\ref{tab:config2t}):
\begin{equation}
\theta_{23} - \frac{1}{4} \, \theta_C^2 \simeq \frac{\pi}{4} \, . \nonumber
\end{equation}
This type of sum rules can be used to falsify our model if {\em all} of the contributing
observables (at least one from the lepton sector) are measured
\item[Type~III sum rules (observable-model sum rules)]
 Sum rules relating  observables from one or both sectors to unphysical quantities or model parameters.
 We distinguish two types:
\begin{description}
\item[Type~IIIa sum rules]
relate observables to unphysical parameters, such as the mixing angles or phases in $U_\ell$ or $U_\nu$ (see, \eg, \Refs~\cite{Masina:2005hf,Antusch:2005kw}). Example (\cf, \Tab~\ref{tab:config2t}):
\begin{equation}
\theta_{12} + \frac{1}{\sqrt{2}} \, \cos \delta^\ell \, \theta_C \simeq \frac{\pi}{4} \, . \nonumber
\end{equation}
\item[Type~IIIb sum rules]
relate observables to model parameters, such as the phase $\Phi$ introduced just before \equ{texture}. Example (\cf, \Tab~\ref{tab:config2t}):
\begin{equation}
\sin \deltacp \simeq \sqrt{2} \, \theta_C \, \sin 2 \Phi \, . \nonumber
\end{equation}
\end{description} 
These sum rules can be used to obtain (model-dependent) information on the unphysical quantities or model parameters if  {\em all} of the contributing
observables (at least one from the lepton sector) are measured.
\end{description}

\subsection{Parameterization of the mixing matrices} 

Let us now give some details
on the parameterization of our mixing matrices. 
A general unitary $3\times 3$ matrix $U_\text{unitary}$ can be written as
\begin{equation}\label{equ:unitary}
 U_\text{unitary}=
 \Gamma \cdot \text{diag}\left(1,e^{\text{i}\widehat{\varphi}_{1}},e^{\text{i} \widehat{\varphi}_{2}}\right)\cdot\widehat{U}\cdot\text{diag}\left(e^{\text{i} \widehat{\alpha}_{1}},e^{\text{i}
 \widehat{\alpha}_{2}}, 1 \right) = \Gamma \cdot D \cdot \widehat{U} \cdot K
 ,
\end{equation}
where the phases $\widehat{\varphi}_1$, $\widehat{\varphi}_2$, $\widehat{\alpha}_1$, and $\widehat{\alpha}_2$, take
their values in the interval $\left[0,2\pi\right]$, $\Gamma$ is a global phase factor, and
$\widehat{U}$ is the (CKM-like) standard parameterization in \equ{ckm}.
We then obtain the PMNS mixing matrix from the charged current interaction Lagrangian 
as
\begin{equation}\label{equ:pmnspara}
U_\text{PMNS}=U_\ell^\dagger U_\nu=K_\ell^\dagger \widehat{U}_\ell^\dagger D_\ell^\dagger D_\nu \widehat{U}_\nu\, K_\nu = \widehat{U}_\ell^\dagger D \widehat{U}_\nu\, K_\nu \, .
\end{equation}
 Note that we have already ignored  an unphysical overall phase in
$U_\text{PMNS}$. Furthermore, as the last step,
we have rotated away $K_\ell$ by re-phasing the charged lepton fields, and we
have defined $D=D_\ell^\dagger D_\nu$. We define the remaining phases by
$D=\text{diag}(1,e^{\text{i} \varphi_1},e^{\text{i} \varphi_2})$
and $K_\nu=\text{diag}(e^{\text{i} \alpha_1},e^{\text{i} \alpha_2},1)$.
The effect of the phases in $D$ has, for example, been discussed in \Ref~\cite{Masina:2005hf}.
 The PMNS matrix in
\equ{pmnspara} can, physics-wise equivalently, be directly written as
\begin{equation}\label{equ:standard}
 U_\text{PMNS}=U_\ell^\dagger U_\nu=\widehat{U}\cdot\text{diag}(e^{\text{i}\phi_1},e^{\text{i}\phi_2},1),
\end{equation}
where $\widehat{U}$ is a CKM-like matrix that is on the form as in \equ{ckm}, and the phases $\phi_{1}$ and $\phi_2$ are the Majorana
phases. The CKM-like matrix $\widehat{U}$ in \equ{standard} is
described by the solar angle $\theta_{12}$, the reactor angle
$\theta_{13}$, the atmospheric angle $\theta_{23}$, and one Dirac
CP-phase $\delta$, which we identify in the standard
parameterization of \equ{ckm} by using
$\hat{\theta}_{ij}\rightarrow\theta_{ij}$ and
$\widehat{\delta}\rightarrow\delta$. 
Note that we have used
$U_\ell = \widehat{U}_\ell$, and that we will use $U_\ell'=D_\ell' \widehat{U}_\ell' K_\ell'=\widehat{U}_\ell' K_\ell'$ in \equ{lepdiag}, \ie,
in all cases, $D_\ell=1$, $K_\ell=1$, and $D_\ell'=1$. Consequently, $D=D_\nu$ in \equ{pmnspara}.\footnote{This restriction neither affects \equ{pmnspara} and its fit to data, nor constrains 
the effective neutrino mass matrix \equ{Maj}. In addition, the constraint $K_\ell=1$ does not 
change \equ{lepdiag} because it only appears a a product with $K_\ell'$. However, the choices $D_\ell=D_\ell'=1$, which are multiplied to \equ{lepdiag} from the outside, limit the set of allowed phases in $M_\ell$. One should keep this additional degree of freedom in mind when one interprets our examples for 
possible $M_\ell$.} With these restrictions, the set of parameters
\begin{equation}
(\theta_{12}^\ell, \theta_{13}^\ell, \theta_{23}^\ell, \delta^\ell, \theta_{12}^\nu, \theta_{13}^\nu, \theta_{23}^\nu,\delta^\nu, \varphi_1, \varphi_2, \alpha_1,\alpha_2)
\label{equ:params}
\end{equation}
fully determines \equt{Maj}{pmnspara}. For our configurations, we will later on fix some of the $\theta_{ij}$'s to $\pi/4$, and assume the other $\theta_{ij}$'s to be small. For the phases, we initially do not make any special assumptions.
If one is interested in the specific form of the charged lepton mass matrix \equ{lepdiag}, one needs to specify the set of parameters
\begin{equation}
(\theta_{12}^{\ell'}, \theta_{13}^{\ell'}, \theta_{23}^{\ell'}, \delta^{\ell'}, \alpha_1^{\ell'}, \alpha_2^{\ell'})
\label{equ:params2}
\end{equation}
as well.

\subsection{Calculation of observables by re-phasing invariants}

As soon as $U_{\mathrm{PMNS}}$ is computed from a set of parameters \equ{params} used in \equ{pmnspara}, one can extract the physical observables by using re-phasing invariants (see, \eg, \Ref~\cite{Jenkins:2007ip}). We choose the three quadratic re-phasing invariants $|U_{13}|^2$, $|U_{23}|^2$ and $|U_{12}|^2$ to determine the three angles $\theta_{13}$, $\theta_{23}$ and $\theta_{12}$ as follows:
\begin{align}
\sin^2\theta_{13}&= |U_{13}|^2\\
\sin^2\theta_{23}&= \frac{|U_{23}|^2}{1-|U_{13}|^2}\\
\sin^2\theta_{12}&= \frac{|U_{12}|^2}{{1-|U_{13}|^2}}.
\end{align}
The observable phases $\deltacp$ and the two Majorana phases $\phi_1$ and $\phi_2$ are determined with the quadratic invariant $|U_{22}|^2$ and these CP-odd quartic invariants: the Jarlskog invariant $J=\mathrm{Im}(U_{22} U_{11} U^*_{21} U^*_{12})$~\cite{Jarlskog:1985ht}, $(U_{13}U_{11}^*)^2$, and $(U_{13}U_{12}^*)^2$. 
The two following equations determine $\deltacp$ and its sign:
\begin{equation}
\begin{split}
J=&c_{12}c_{23}c_{13}^2s_{23}s_{13}s_{12}\sin{\deltacp} \, ,  \\
|U_{22}|^2=&c_{12}^2c_{23}^2+s_{12}^2s_{23}^2s_{13}^2-c_{12}c_{23}s_{12}s_{23}s_{13}2\cos{\deltacp} \, .
\end{split}
\end{equation}
Since in our definition of the Majorana phases (in \equ{standard}) physical quantities depend only on $2\phi_1$ and $2\phi_2$, we only determine their double values as
\begin{equation}
\begin{split}
2\phi_1=&-\arg((U_{13}U_{11}^*)^2)-2\deltacp \, , \\
2\phi_2=&-\arg((U_{13}U_{12}^*)^2)-2\deltacp  \, .
\end{split}
\end{equation}

\subsection{Consistency check with data}

As the last step, we will compare the extracted observables to data in order to check potential constraints on the unphysical quantities. Whenever we make quantitative estimates, we will use the following numbers~\cite{Schwetz:2006dh}: 
\begin{align}
\sin^2 \theta_{13}  & \lesssim   0.04 \quad (3\sigma) \, , \nonumber \\
\sin^2 \theta_{12} & = 0.3 \quad \pm \, \, \,  9  \%  \quad (1\sigma) \, , \quad 0.24 \lesssim \sin^2 \theta_{12} \lesssim 0.40 \quad (3\sigma) \, , \nonumber \\
\sin^2 \theta_{23} & = 0.5 \quad \pm  16 \%  \quad (1\sigma) \, , \quad 0.34 \lesssim \sin^2 \theta_{23} \lesssim 0.68 \quad (3\sigma)  \, .
\label{equ:constraints}
\end{align}
We choose the $3\sigma$ allowed ranges, unless explicitely mentioned otherwise.

\section{Entangled maximal mixing configurations}
\label{sec:obs}

In this section, we discuss two different configurations with maximal mixing angles in both $U_\ell$ and $U_\nu$, 
where the stars denote small mixing angles $\mathcal{O}(\theta_C)$: 
\begin{enumerate}
\item
Three maximal mixing angles: $(\theta_{12}^\ell, \theta_{13}^\ell, \theta_{23}^\ell,
\theta_{12}^\nu, \theta_{13}^\nu, \theta_{23}^\nu) = (*,\frac{\pi}{4}, \frac{\pi}{4}, *, \frac{\pi}{4}, *)$
 \item
Two maximal mixing angles: $(\theta_{12}^\ell, \theta_{13}^\ell, \theta_{23}^\ell, \theta_{12}^\nu, \theta_{13}^\nu, \theta_{23}^\nu)=(*,*, \frac{\pi}{4}, \frac{\pi}{4}, *, *)$
\end{enumerate}
Since the small mixing angles are $\mathcal{O}(\theta_C)$, we can use them for expansions.
Note that these configurations are not the only ones leading to a valid $U_{\mathrm{PMNS}}$.
We show a list of possible configurations in \App~\ref{app:allconfigs}, where we also describe the selection criteria for the ones discussed analytically here. Out of the 44 configurations found, 38 have entangled maximal mixings. Therefore, the cases discussed in this study might be rather typical than the remaining six configurations with maximal mixings in either $U_\ell$ or $U_\nu$ only. 

As described in \Sec~\ref{sec:method}, we compute the observables using invariants, and we discuss the effects from the current bounds. Where applicable, we will discuss different implementations for the small mixing angles.
Note that, at this point, we do not assume specific values for the small mixing angles, such as powers of $\theta_C$. Therefore, there will not be any connection to the quark sector yet. However, our conclusions will be less model-dependent than the ones from the following section. 

\subsection{Three maximal mixing angles: $\boldsymbol{(\theta_{12}^\ell, \theta_{13}^\ell, \theta_{23}^\ell,
\theta_{12}^\nu, \theta_{13}^\nu, \theta_{23}^\nu) = (*,\frac{\pi}{4}, \frac{\pi}{4}, *, \frac{\pi}{4}, *)}$}
\label{sec:configuration1}

This configuration is given by three maximal mixing angles, one in the neutrino sector, $\theta^\nu_{13}$, and two in the charged lepton sector, $\theta^\ell_{13}$ and $\theta^\ell_{23}$. Note again that the other mixings angles $\theta^\ell_{12}$, $\theta^\nu_{12}$, and $\theta^\nu_{23}$ are assumed to be small, \ie,  $\mathcal{O}(\theta_C)$, which means that we can use them for expansions.
For the sake of simplicity, it is convenient to use some phase re-definitions for this configuration:
\begin{equation}
\chi \equiv\delta^\ell-\varphi_1 \, , \quad 
\beta \equiv\delta^\ell-\delta^\nu-\varphi_2 \, , \quad
\gamma \equiv\varphi_1-\varphi_2 \, .
\label{equ:beta}
\end{equation}
With these re-definitions, we expand $\sin^2\theta_{13}$ to first order as\footnote{In the following, we will expand
the observables to the minimum order for which we recover the main qualitative features. For example, if the zero{\it th} order vanishes, or the terms from different orders are numerically comparable, we will include higher orders.}  
\begin{equation}
\label{equ:s13}
\begin{split}
\sin^2\theta_{13}=\frac{3}{8}-\frac{\sqrt{2}}{4} \cos{\beta} \ +\ & \theta^\ell_{12} \left(\frac{1}{2} \cos{(\delta^\ell-\beta)}-\frac{\sqrt{2}}{4} \cos{\delta^\ell}\right) + \\
 +\ & \theta^\nu_{23} \left(\frac{1}{4} \cos{\gamma}-\frac{\sqrt{2}}{4} \cos{(\gamma-\beta)}\right)+\mathcal{O}(\theta_{k}^2) \, .
\end{split}
\end{equation}
Here the symbol $\theta_k^2$ stands for all small mixing angles and combinations thereof, such as $\theta^\ell_{12}\theta^\nu_{23}$ and $(\theta^\nu_{23})^2$. 
Obviously, the zero{\it th} order term in this expansion is given by $3/8-\sqrt{2}\cos{\beta}/4$, which means that $\sin^2 \theta_{13}$ is expected to be large. From \equ{constraints}, however, we know that $\sin^2\theta_{13}\lesssim0.04$, which leads to  $\cos{\beta}\gtrsim0.95$ to zero{\it th} order.

The atmospheric mixing angle is given by
 \begin{equation}
\label{equ:s23}
\begin{split}
\sin^2\theta_{23}=&\frac{2}{5+2\sqrt{2}\cos{\beta}} \\
 +& \frac{2}{(5+2\sqrt{2}\cos{\beta})^2}\theta^\ell_{12} \left(-4\cos{(\delta^\ell-\beta)}+\sqrt{2}\cos{\delta^\ell}+2\cos{(\delta^\ell+\beta)}-2\sqrt{2}\cos{(2\beta-\delta^\ell)}\right)-\\
 -& \frac{4}{(5+2\sqrt{2}\cos{\beta})^2}\theta^\nu_{23} \left(4\cos{\gamma}+2\sqrt{2}\cos{(\gamma-\beta)}+\sqrt{2}\cos{(\beta+\gamma)}\right)+\mathcal{O}(\theta_{k}^2).
\end{split}
\end{equation}
This implies that to zero{\it th} order, we have $\sin^2\theta_{23}=2/(5+2\sqrt{2}\cos{\beta})$.
Since we have a lower bound for $\cos{\beta}$ from the $\sin^2 \theta_{13}$ upper bound,  $\sin^2\theta_{23}$ 
must be smaller than maximal mixing. From the zero{\it th} order constraint on $\cos{\beta}$, we obtain  $\sin^2\theta_{23}\lesssim0.26$ to zero{\it th} order, which is below the allowed range for $\sin^2\theta_{23}$ in \equ{constraints}. Therefore, one has to include higher order corrections such that $\sin^2\theta_{23}$ is shifted up into the allowed range. Nevertheless, it will stay on the lower edge of the allowed range, which makes this configuration easily testable in future experiments.

As the third observable, the solar mixing angle is determined by
\begin{equation}
\label{equ:s12}
\begin{split}
\sin^2\theta_{12}=&\frac{2}{5+2\sqrt{2}\cos{\beta}} \\
 +& \frac{8}{(5+2\sqrt{2}\cos{\beta})^2}\theta^\ell_{12} \left(2\cos{(\delta^\ell-\beta)}+2\sqrt{2}\cos{\delta^\ell}+\cos{(\delta^\ell+\beta)}\right)-\\
 -& \frac{4}{(5+2\sqrt{2}\cos{\beta})^2}\theta^\nu_{23} \left(4\cos{\gamma}+2\sqrt{2}\cos{(\gamma-\beta)}+\sqrt{2}\cos{(\beta+\gamma)}\right)-\\
 -& \frac{2}{5+2\sqrt{2}\cos{\beta}} \theta^\nu_{12}\left(2\cos{\chi}+\sqrt{2}\cos{(\chi-\beta)}\right)+\mathcal{O}(\theta_{k}^2)\\
\end{split}
\end{equation}
To zero{\it th} order, $\sin^2\theta_{12}=2/(5+2\sqrt{2}\cos{\beta}) = \sin^2 \theta_{23}$. Together with the $\cos \beta \gtrsim 0.95$ constraint to zero{\it th} order, this implies that $\sin^2\theta_{12} \simeq 2/(5+2\sqrt{2})\simeq 0.26$  coincides very well with its current best-fit value. 

For the sake of simplicity, we just show the zero{\it th} order contribution for $\sin{\deltacp}$ and $\cos{\deltacp}$:
\begin{eqnarray}
\label{equ:sindelta}
\sin{\deltacp} & = & -\frac{(5+2\sqrt{2}\cos{\beta})\sin{\beta}}{\sqrt{(-3-2\sqrt{2}\cos{\beta})(-5+4\cos{2\beta})}}+\mathcal{O}(\theta_k) \\
\label{equ:cosdelta} \cos{\deltacp} & = & \frac{\sqrt{2}\cos{\beta}+2\cos{2\beta}}{\sqrt{6-4\sqrt{2}\cos{\beta}}(3+2\sqrt{2}\cos{\beta})}+\mathcal{O}(\theta_k).
\end{eqnarray}
Because $\cos{\beta}$ has to be large (and $|\beta|$ therefore has to be small), $|\sin \deltacp|$ is small and $\cos{\deltacp}$ is positive, \ie, $\deltacp$ is close to zero. Furthermore,
$\sin{\deltacp}$ has always the opposite sign of $\sin{\beta}$. To zero{\it th} order in the small mixing angles, $|\deltacp|\lesssim\pi/3$, which means that maximal CP violation requires higher order corrections.

In summary, configuration~1 is characterized by large $\sin^2 \theta_{13}$, strong deviations from maximal atmospheric mixing into the first octant, $\sin^2 \theta_{12}$ close to its current best-fit value, and $\deltacp$ close to $0$. Therefore, future improved bounds on $\sin^2 \theta_{13}$ and an improved precision for $\sin^2 \theta_{23}$ can easily test this configuration. Note that this configuration does not look compatible with the tri-bimaximal mixing idea since $\sin^2 \theta_{13}$ is intrinsically large and $\sin^2 \theta_{23}$ strongly deviates from maximal mixing.
 Our procedure has been particularly simple in this case: We have  expanded the observables in the small mixing angles, and we have discussed the implications of the current bounds on the unphysical quantities and observables. For example, we have found from the $\sin^2 \theta_{13}$ bound that $\cos \beta=\cos( \delta^\ell-\delta^\nu-\phi_2)$ has to be close to one, and we have used this knowledge for the other observables. The next example will be more sophisticated in the sense that there are more qualitative cases.

\subsection{Two maximal mixing angles: $\boldsymbol{(\theta_{12}^\ell, \theta_{13}^\ell, \theta_{23}^\ell, \theta_{12}^\nu, \theta_{13}^\nu, \theta_{23}^\nu)=(*,*, \frac{\pi}{4}, \frac{\pi}{4}, *, *)}$}
\label{sec:configuration2}

This configuration~2 is characterized by one large mixing in each the neutrino and charged lepton sectors, namely $\theta^\ell_{23}$ and $\theta^\nu_{12}$. The small mixing angles $\theta^\ell_{12}$, $\theta^\ell_{13}$, $\theta^\nu_{13}$ and $\theta^\nu_{23}$ are, except from $\sin^2\theta_{13}$, expanded up to second order. Since there is one more small mixing angle than before, there is naturally more freedom involved in this case.

Again, we start the discussion of the observables with $\sin^2\theta_{13}$, which we expand to third order as\footnote{We expand $\sin^2 \theta_{13}$ to the third order in the small mixing angles because the leading contributions come from the second order terms, and the coefficients of the third order terms are comparatively large.}
\begin{equation}
\label{equ:s132}
\begin{split}
\sin^2 \theta_{13}=&\frac{1}{2}\ \theta^{\ell\ 2}_{12}+\frac{1}{2}\ \theta^{\ell\ 2}_{13}+\theta^{\nu\ 2}_{13}-\cos{\delta^\ell}\ \theta^\ell_{12} \theta^\ell_{13}+\sqrt{2}\cos{(\varphi_2+\delta^\nu)}\ \theta^\ell_{12}\theta^\nu_{13}\\
&-\sqrt{2}\cos{(\varphi_2-\delta^\ell+\delta^\nu)}\ \theta^\ell_{13}\theta^\nu_{13}-6\cos{(\varphi_1-\varphi_2)}\ \theta^{\ell\ 2}_{12}\theta^\nu_{23})\\
&-12 \sin{\delta^\ell}\sin{(\varphi_1-\varphi_2)}\ \theta^\ell_{12}\theta^\ell_{13}\theta^\nu_{23}+6\cos{(\varphi_1-\varphi_2)}\ \theta^{\ell\ 2}_{13}\theta^\nu_{23}\\
&-6\sqrt{2}\cos{(\delta^\nu+\varphi_1)}\ \theta^\ell_{12}\theta^\nu_{13}\theta^\nu_{23}-6\sqrt{2}\cos{(\delta^\nu+\varphi_1-\delta^\ell)}\ \theta^\ell_{13}\theta^\nu_{13}\theta^\nu_{23}+\mathcal{O}(\theta_{k}^4) \, . 
\end{split}
\end{equation}
Since  $\sin^2 \theta_{13}$ is zero to zero{\it th} and first order, it is intrinsically small compared to that of configuration~1. However, if the small angles are chosen comparatively large, it is possible to generate a $\sin^2 \theta_{13}$ close to the current bound.
There is already one important qualitative difference observable at this place, which is different from configuration~1: If all small mixing angles are zero, $\deltacp$ is not defined. This means that we will need to distinguish different cases determined by the non-vanishing small mixing angles.

The solar mixing angle is given by
\begin{equation}
\label{equ:s122}
\begin{split}
\sin^2\theta_{12}=&\frac{1}{2}-\frac{1}{\sqrt{2}}\cos{(\delta^\ell-\varphi_1)}\ \theta^\ell_{13}-\frac{1}{\sqrt{2}}\cos{\varphi_1}\ \theta^\ell_{12}\\
&-\frac{1}{\sqrt{2}}\cos{\varphi_2}\ \theta^\ell_{12}\theta^\nu_{23}+\frac{1}{\sqrt{2}}\cos{(\delta^\ell-\varphi_2)}\ \theta^\ell_{13}\theta^\nu_{23}+\mathcal{O}(\theta_{k}^3) \, ,
\end{split}
\end{equation} 
which is maximal to zero{\it th} order. Therefore,
in order to satisfy the bounds on $\sin^2\theta_{12}$ in \equ{constraints}, at least one of the two small charged lepton angles $\theta^\ell_{12}$ or $\theta^\ell_{13}$ has to be moderately large. Nevertheless, $\sin^2 \theta_{12}$ close to its best-fit value will require some tuning of the phases.
Note that once $\theta^\ell_{12}$ or $\theta^\ell_{13}$ is non-vanishing, an exactly vanishing $\sin^2 \theta_{13}$ in \equ{s132} will also require some tuning.

We expand $\sin^2 \theta_{23}$ to second order in the small angles:
\begin{equation}
\label{equ:s232}
\begin{split}
\sin^2\theta_{23}=&\frac{1}{2}-\cos{(\varphi_1-\varphi_2)}\ \theta^\nu_{23}-\frac{1}{4}\ \theta^{\ell\ 2}_{12}+\frac{1}{4}\ \theta^{\ell\ 2}_{13}+\frac{1}{2}\cos{\delta^\ell}\ \theta^\ell_{13}\theta^\ell_{12}\\
&-\frac{1}{\sqrt{2}}\cos{(\delta^\ell-\delta^\nu-\varphi_2)}\ \theta^\ell_{13}\theta^\nu_{13}-\frac{1}{\sqrt{2}}\cos{(\delta^\nu+\varphi_2)}\ \theta^\ell_{12}\theta^\nu_{13}+\mathcal{O}(\theta_{k}^3).
\end{split}
\end{equation}
Obviously, $\sin^2 \theta_{23}$ is to zero{\it th} order given by its best-fit value maximal mixing, and $\sin^2\theta_{12}$ and $\sin^2\theta_{23}$ are identical to zero{\it th} order. Therefore, we do not obtain additional constraints from \equ{constraints}. However, we will see that the higher order contributions lead to interesting observations for deviations from maximal mixing and the $\theta_{23}$ octant.

As indicated above, we cannot calculate $\deltacp$ as a expansion in the small angles without further assumptions. If all of the contributing small angles are zero, $\sin^2 \theta_{13}$ will be zero, and hence $\deltacp$ will not be defined. In addition, at least one of the two angles $\theta^\ell_{12}$ and $\theta^\ell_{13}$ has to be sufficiently large to move $\sin^2\theta_{12}$ in the currently allowed range. Therefore, we choose three different sets of small angles: all small angles zero except $\theta^\ell_{12}$, all small angles zero except $\theta^\ell_{13}$, and $\theta^\ell_{12}=\theta^\ell_{13}=\sqrt{\theta^\nu_{23}}$ with $\theta^\nu_{13}=0$. The third case will allow us to construct tri-bimaximal mixing. In the following, we expand the observable mixing angles to second order and the observable Dirac CP phase to first order.

\subsubsection{Non-vanishing $\boldsymbol{\theta_{12}^\ell}$, and $\boldsymbol{\theta_{13}^\ell=\theta_{13}^\nu=\theta_{23}^\nu = 0}$}
\label{sec:t130}

In this case, we have
\begin{align}
\label{equ:s132t2} 
\sin^2\theta_{13}=&\frac{1}{2}(\theta^\ell_{12})^2 \, ,\\
\label{equ:s122t2} 
\sin^2\theta_{12}=&\frac{1}{2}-\frac{1}{\sqrt{2}}\cos{\varphi_1} \, \theta^\ell_{12} \, ,\\
\label{equ:s232t2} 
\sin^2\theta_{23}=&\frac{1}{2}-\frac{1}{4}(\theta^\ell_{12})^2 \, ,\\
\label{equ:delta2t2} 
\sin{\deltacp}=& \sin{\varphi_1} \, , \quad \cos{\deltacp}=-\cos{\varphi_1} \, .
\end{align}
In \equ{s122t2}, we have to choose the product $\cos{\varphi_1} \, \theta^\ell_{12}$ positive and moderately large in order to be within the allowed range for $\sin^2\theta_{12}$. This implies a significant deviation of $\sin^2\theta_{13}$ from zero, and of $\sin^2\theta_{23}$ from maximal mixing. Note that $\deltacp$ is given by $\deltacp=\pi-\varphi_1$, and that $\cos \varphi_1$ is constrained by the bound on $\sin^2\theta_{12}$ (\cf, \equ{constraints}) by $\cos{\varphi_1}\gtrsim\sqrt{2}/(10 \, \theta^\ell_{12})$. Therefore, $\deltacp$ tends to be close to $\pi$, and maximal CP violation is forbidden because of $\cos{\varphi_1} > 0$. Similarly, from $\theta_{12}^\ell \gtrsim \sqrt{2}/(10 \cos \varphi_1) \ge \sqrt{2}/10$, we obtain a lower bound $\sin^2 \theta_{13} \gtrsim 0.01$. This bound lies within the range of upcoming reactor experiments, such as Double Chooz (see, \eg, \Refs~\cite{Huber:2006vr,Ardellier:2006mn}).

As the next step, we can eliminate $\theta^\ell_{12}$ in order to establish relationships among the physical observables. Such relationships are {\bf type~I sum rules} according to our definition:
\begin{align}
\label{equ:srtheta23t2}
 \theta_{23}+\frac{1}{2}\,\theta_{13}^2&\simeq\frac{\pi}{4} \, , \\
\label{equ:srtheta12t2} 
\theta_{12}-\cos{\deltacp}\,\theta_{13}&\simeq\frac{\pi}{4} \, .
\end{align}
The deviation from maximal mixing described by \equ{srtheta23t2}, which can be as large as $3\%$ from the $\sin^2 \theta_{13}$ upper bound in \equ{constraints}, is on the edge of future experiments (see, \eg, \Ref~\cite{Antusch:2004yx}).
 In \equ{srtheta12t2}, the observables $\theta_{12}$, $\theta_{13}$, and $\deltacp$ are related. Obviously, 
this case will be challenged for either better $\theta_{13}$ bounds, or a detection of $\theta_{13}>0$ together with $\cos \deltacp >0$.

\subsubsection{Non-vanishing $\boldsymbol{\theta_{13}^\ell}$, and $\boldsymbol{\theta_{12}^\ell=\theta_{13}^\nu=\theta_{23}^\nu = 0}$}
\label{sec:t120}

Here we find
\begin{align}
\label{equ:s132t3}
 \sin^2\theta_{13}=&\frac{1}{2}(\theta^\ell_{13})^2 \, ,\\
\label{equ:s122t3} 
\sin^2\theta_{12}=&\frac{1}{2}-\frac{1}{\sqrt{2}}\cos{(\delta^\ell-\varphi_1)} \, \theta^\ell_{13} \, ,\\
\label{equ:s232t3} 
\sin^2\theta_{23}=&\frac{1}{2}+\frac{1}{4}(\theta^\ell_{13})^2 \, ,\\
 \sin{\deltacp}=&\sin{(\delta^\ell-\varphi_1)} \, , \quad \cos{\deltacp}=\cos{(\delta^\ell-\varphi_1)} \, .
\end{align}
Obviously, we recover the same structure as in the previous case  with $\theta^\ell_{13} \leftrightarrow  \theta^\ell_{12}$ and $\delta^\ell-\varphi_1 \leftrightarrow \varphi_1$. 
The main differences can be easily seen in the type~I sum rules
\begin{align}
\label{equ:srtheta23t3}
 \theta_{23}-\frac{1}{2}\ \theta_{13}^2&\simeq\frac{\pi}{4} \, ,\\
\label{equ:srtheta12t3} 
\theta_{12}+\cos{\deltacp}\,\theta_{13}&\simeq\frac{\pi}{4} \, .
\end{align}
Compared to \equ{srtheta23t2}, $\theta_{23}$ is now in the second octant, \ie, the deviation from maximal mixing is positive instead of negative. The difference between these two cases 
should now be measurable by future experiments, because it is twice as big as the deviation from maximal mixing. In addition, this comparison motivates a precision octant measurement, which could discriminate which of $\theta_{12}^\ell$ and $\theta_{13}^\ell$ dominates in this framework. Compared to \equ{srtheta12t2}, $\deltacp$ is now located at around $0$ instead of $\pi$. This means that even if no CP violation is found in future experiments, it is interesting to distinguish these CP-conserving values (see, \eg, \Ref~\cite{Huber:2004gg} for precision measurements of $\deltacp$).

\subsubsection{Large CP violation or tri-bimaximal mixing}
\label{sec:tribima}

Here we discuss the case $X \equiv \theta^\ell_{12}=\theta^\ell_{13}=\sqrt{\theta^\nu_{23}}$ and $\theta_{13}^\nu=0$,
\ie, $(\theta_{12}^\ell, \theta_{13}^\ell, \theta_{23}^\ell,
\theta_{12}^\nu, \theta_{13}^\nu, \theta_{23}^\nu) = (X,X, \frac{\pi}{4}, \frac{\pi}{4}, 0, X^2)$, in order to construct maximal CP violation or the tri-bimaximal case. For the observables, we find
\begin{align}
\label{equ:s132x}
\sin^2\theta_{13}=&X^2\left(1-\cos{\delta^\ell}\right) \, ,\\
\label{equ:s122x} 
\sin^2\theta_{12}=&\frac{1}{2}-\frac{1}{\sqrt{2}}\,X\left(\cos{(\delta^\ell-\varphi_1)}+\cos{\varphi_1}\right) \, , \\
\label{equ:s232x} 
\sin^2\theta_{23}=&\frac{1}{2}\left(1+\left(\cos{\delta^\ell}+2\cos{(\varphi_1-\varphi_2)}\right)X^2\right) \, ,\\ 
\label{equ:delta2x}
\sin{\deltacp}=&\frac{\sin{(\delta^\ell-\varphi_1)}+\sin{\varphi_1}}{\sqrt{2-2\cos{\delta^\ell}}} \, , \quad
 \cos{\deltacp}=\frac{\cos{(\delta^\ell-\varphi_1)}-\cos{\varphi_1}}{\sqrt{2-2\cos{\delta^\ell}}} \, .
\end{align}
In comparison to the previous cases, we can have a vanishing $\sin^2\theta_{13}$ here. Furthermore, $\sin^2\theta_{13}$ and $\sin^2\theta_{23}$ depend on the phase $\delta^\ell$. Except from that, the structure of the equations is similar to those of \Secs~\ref{sec:t130} and~\ref{sec:t120}. For $X$, we obtain from Eqs.~(\ref{equ:constraints}) and~(\ref{equ:s122x}) the constraint $X \gtrsim \sqrt{2}/20 \simeq 0.07$.

Let us now focus on the possibility of maximal CP violation, which was excluded in the previous cases. For maximal CP violation, $\cos{\deltacp}=0$, and we obtain from \equ{delta2x} $\tan{\varphi_1}=(1-\cos{\delta^\ell})/\sin{\delta^\ell}$. Using this condition in \equ{s132x} and \equ{s122x}, one can show that  $\sin^2\theta_{12}$ and $\sin^2\theta_{13}$ are compatible with their currently allowed ranges. For CP conservation, however, we obtain from $\sin{\deltacp}=0$ in \equ{delta2x} the condition $\tan{\varphi_1}=\sin{\delta^\ell}/(\cos{\delta^\ell}-1)$, which is incompatible with the  $\sin^2\theta_{12}$ allowed range. Therefore, CP conservation is excluded, and maximal CP violation is possible. 
More specifically, one can show that $|\sin \deltacp| \gtrsim 0.4$  for $X \le \theta_C$, \ie, $|\deltacp| \gtrsim \pi/8$ for $\cos \deltacp>0$. This implies that a future experiment will find the CP violation if the fraction of $\deltacp$, for which CP violation can be discovered, is about 75\%. A neutrino factory would find such a CP violation for $\stheta \gtrsim 10^{-4}$ (see, \eg, \fig~23 in \Ref~\cite{Huber:2006wb}). 

\begin{table}[t]
\begin{center}
\small{
\begin{tabular}{l|lll}
\hline
& \multicolumn{3}{c}{Mixing angle case $(\theta_{12}^\ell, \theta_{13}^\ell, \theta_{23}^\ell,
\theta_{12}^\nu, \theta_{13}^\nu, \theta_{23}^\nu)$ } \\
Observable &$(Y, 0, \frac{\pi}{4}, \frac{\pi}{4}, 0, 0)$&$(0, Y, \frac{\pi}{4}, \frac{\pi}{4}, 0, 0)$& $(X,X, \frac{\pi}{4}, \frac{\pi}{4}, 0, X^2)$\\
\hline
$\theta_{12} $&$\simeq \pi/4+\cos{\deltacp} \, \theta_{13}$&$\simeq \pi/4-\cos{\deltacp} \, \theta_{13}$&
Any in currently allowed range  \\
$\theta_{23}$&$\simeq\pi/4-\theta_{13}^2/2$&$\simeq\pi/4+\theta_{13}^2/2$&  Around maximal mixing \\
$\theta_{13}$& $\gtrsim 0.1$ & $\gtrsim 0.1$ & Any in currently allowed range \\ 
$\deltacp$&$\sim \pi$&$\sim 0$& $|\sin \deltacp| \gtrsim 0.4$ (if $\theta_{13}>0$) \\
\hline
\end{tabular}
} 
\end{center}
\mycaption{\label{tab:config2} Type~I sum rules or approximative values/ranges for the observables $\deltacp$, $\theta_{13}$, $\theta_{23}$ and $\theta_{12}$ for the three cases cases of configuration~2 (\cf, \Secs~\ref{sec:t130}, \ref{sec:t130}, and \ref{sec:tribima}). The angle $Y$ refers to a mixing angle in the range $\sqrt{2}/10 \lesssim Y \lesssim \theta_C$, whereas the angle $X$ lies within the range $\sqrt{2}/20 \lesssim X \lesssim \theta_C$. The conditions for the last column are much weaker than in the other two cases,
because the dependence on three unphysical phases allows for more freedom. 
}
\end{table}

We can construct tri-bimaximal mixing with this set of assumptions. For exactly tri-bimaximal mixing, the following equations have to be fulfilled: $\sin^2\theta_{12}=1/3$, $\sin^2\theta_{13}=0$, and $\sin^2\theta_{23}=1/2$. From \equ{s132x}, we first of all obtain $\delta^\ell=0$. Therefore, \equt{s122x}{s232x} simplify to
\begin{align}
 \sin^2\theta_{12}=&\frac{1}{2}-\sqrt{2}\cos{\varphi_1}X\stackrel{!}{=}\frac{1}{3} \, \\
\sin^2\theta_{23}=&\frac{1}{2}+X^2\left(\frac{1}{2}-\cos{(\varphi_1-\varphi_2)}\right)\stackrel{!}{=}\frac{1}{2} \, .
\end{align}
Eliminating $\varphi_1$, the condition for tri-bimaximal mixing is then given in terms of $\varphi_2$ and $X$ as
\begin{equation}
 \cos{(\frac{\pi}{3}\pm\varphi_2)}\,X=\frac{1}{6\sqrt{2}} \, .
\end{equation}
This condition can be fulfilled, for instance, for $\varphi_2=\pi/3$ and $X=1/(6\sqrt{2})$. 
This means that tri-bimaximal mixing can be constructed within configuration~2. Note that we have only expanded the small mixing angles to second order, which means that higher order corrections will act as small deviations from tri-bimaximal mixing. In addition, note that $\deltacp$ is, of course, not defined if $\sin^2 \theta_{13}$ is exactly zero.

We summarize the three special cases discussed for configuration~2 in \Tab~\ref{tab:config2}. These three cases can be easily distinguished: If $\sin^2 \theta_{13}$ is much smaller than $0.01$, we have the third case. If $\sin^2 \theta_{13}$ is large, we can use $\deltacp$ to disentangle all three cases. In addition, the $\theta_{23}$ octant could be used to discriminate between the first and second case. 

\section{Introducing mass matrix structure}
\label{sec:matrix}

\begin{table}[t!]
\begin{center}
\small{
\begin{tabular}{llc|cp{7.5cm}}
\hline
&  $(\theta_{12}^\ell, \theta_{13}^\ell, \theta_{23}^\ell,$& &  & $(\theta_{12}^\ell, \theta_{13}^\ell, \theta_{23}^\ell, \delta^\ell, \theta_{12}^\nu, \theta_{13}^\nu, \theta_{23}^\nu,\delta^\nu, \varphi_1, \varphi_2, \alpha_1,\alpha_2)$ \\
\# &   $\theta_{12}^\nu, \theta_{13}^\nu, \theta_{23}^\nu)$ & $T_\nu^{\mathrm{Maj}}$ &  $\Phi$ & Obs.: $(s_{12}^2, s_{13}^2, s_{23}^2, \deltacp, \phi_1, \phi_2)$  \\
\hline
1a & $(*,\frac{\pi}{4}, \frac{\pi}{4}, *, \frac{\pi}{4}, *)$ & $\left(
\begin{array}{ccc}
1 & 0 & 1 \\
0 & \eta & 0 \\
1 & 0 & 1 \\
\end{array}
\right) $ & $\frac{\pi}{2}$ & $(\epsilon^2, \frac{\pi}{4}, \frac{\pi}{4}, 0, 0, \frac{\pi}{4}, 0, 0, 0, 0, 0, 0.79)$ \newline Obs.: $(0.29,0.03,0.25,0,0,3.9)$ \\[0.2cm]
1b &  & $\left(
\begin{array}{ccc}
1 & \eta & 1 \\
\eta & \eta & \eta \\
1 & \eta & 1 \\
\end{array}
\right) $   & $\frac{\pi}{2}$ & $(0, \frac{\pi}{4}, \frac{\pi}{4}, 0, \epsilon, \frac{\pi}{4}, \epsilon, 0, 1.57, 0, 0, 2.4)$  \newline Obs.:  $(0.30,0.03,0.26,1.0,5.9,0.70)$\\[0.2cm]
& & &  $\frac{2\pi}{3}$ & $(0, \frac{\pi}{4}, \frac{\pi}{4}, 0, \epsilon^2, \frac{\pi}{4}, \epsilon, 3.1, 5.2, 3.1, 0, 2.1)$  \newline Obs.:  $(0.29,0.04,0.31,0.39,3.1,1.1)$ \\[0.5cm]
2a &  $(*,*, \frac{\pi}{4}, \frac{\pi}{4}, *, *)$ & $\left(
\begin{array}{ccc}
\eta & \eta & \eta^2 \\
\eta & \eta & 0 \\
\eta^2 & 0 & 1 \\
\end{array}
\right) $  & $\frac{\pi}{2}$ & $(\epsilon, 0, \frac{\pi}{4} , 0, \frac{\pi}{4}, \epsilon^2, 0, 0, 0,  3.1, 0, 0.79)$ \newline Obs.:  $(0.36,0.01,0.50,3.1,0,0.79)$ \\[0.2cm]
2b & & $\left(
\begin{array}{ccc}
\eta & \eta & \eta \\
\eta & \eta & \eta^2 \\
\eta & \eta^2 & 1 \\
\end{array}
\right) $  &  $\frac{\pi}{2}$& $(\epsilon, \epsilon, \frac{\pi}{4} , 4.7, \frac{\pi}{4}, \epsilon, 0, 1.6, 0,  3.1, 0, 0.79)$ \newline Obs.:  $(0.36,0.02,0.47,2.8,0.15,0.65)$ \\[0.2cm]
\hline
\end{tabular}
} 
\end{center}
\mycaption{\label{tab:textures} 
Possible textures for our two configurations if the small mixing angles (``*'') are chosen from the set $\{\epsilon, \epsilon^2, 0 \}$  with $\epsilon = 0.2 \simeq  \theta_C$. Note that one configuration may allow for different textures depending on the choices of the small mixing angles. The parameter $\eta = | \eta | \cdot \exp(i \Phi) = \epsilon \cdot \exp(i \Phi)$. The first column \# refers to the texture number, the second to the configuration, and the third to the texture $T_\nu^{\mathrm{Maj}}$.  The last two columns represent a possible implementation for each texture, \ie, specific choices of $\Phi$, the mixing angles and phases. This implementation includes the parameters used in \equ{pmnspara} (first row), we well as the corresponding observables (second row, where $s_{ij}=\sin \theta_{ij}$). Note that texture (or angle) zeros correspond to $\mathcal{O}(\epsilon^3)$ in this table.}
\end{table}

In the preceding section, we have not made any special assumptions for the small mixing angles in our configurations. In this section, we extend the preceding discussion by two additional aspects:
\begin{itemize}
\item 
 We assume that the small mixings angles $\theta_{ij}^\nu$, $\theta_{ij}^\ell$ in our configurations, which we have denoted by stars, can only come from the set $\{\epsilon, \epsilon^2, 0 \}$, where $\epsilon \simeq \theta_C$.
\item
 We discuss our configurations together with specific mass textures of the form of \equ{texture} for $M^\mathrm{Maj}_\nu$. This means that we include the form of specific mass matrices in the discussion, and that the individual texture entries can only  be 
\begin{equation}
T_{ij} = \eta^{n_{ij}} = \epsilon^{n_{ij}} \, \exp(i n_{ij} \Phi) 
\label{equ:eqlc}
\end{equation} 
with $n_{ij} \in \{0,1,2, \hdots\}$ and an arbitrary but fixed phase $\Phi$.
\end{itemize}
The first aspect establishes, similar to EQLC, a phenomenological connection to the quark sector. Therefore, we will obtain sum rules including the Cabibbo angle. The second aspect will constrain the physical observables and, especially, the Dirac and Majorana phases. In many cases, we will derive a relationship to the phase $\Phi$. Of course, both of these aspects introduce more model-dependence than in the preceding section. However, we will also obtain stronger constraints from the textures. 

We list the textures which we will study in this section together with the corresponding configurations in \Tab~\ref{tab:textures}. These textures have been found in \Ref~\cite{Winter:2007yi} as valid patters which can be fit to data by appropriate choices of the order one coefficients.  We give in the last column one possible set of parameters corresponding to \equ{params}, which allows for a full construction of $M_\nu^{\mathrm{Maj}}$ including order one coefficients for the $\Phi$ given in the second-last column (\cf, \Sec~\ref{sec:method}). In addition, for this set of parameters, the observables  are given in the last column (second rows). The effective 3$\times$3 Majorana texture is computed from \equ{Maj} assuming a normal neutrino mass hierarchy of the form $m_1:m_2:m_3=\epsilon^2:\epsilon:1$ (\cf, \equ{numassratios}).  Note that for one configuration, there can be more possible textures, depending on the choices of the small mixing angles. Similarly, for one texture, there can be many valid sets of parameters, depending on the small mixing angles and phases used. 

\begin{table}[t!]
\begin{center}
\begin{tabular}[t]{c|ll|ll}
\hline
Cond. &Texture 1a&Texture 1b&Texture 2a&Texture 2b\\
\hline
  1&$\varphi_2+\delta^\nu=0$& $\varphi_2+\delta^\nu=0$&$\varphi_1=0$&$\varphi_1=0$ \\
  2&$2(\alpha_2+\varphi_1+\delta^\nu)=\Phi$&$2(\alpha_2+\varphi_1+\delta^\nu)=\Phi$&$2(\alpha_2-\varphi_2)=\Phi$&$2(\alpha_2-\varphi_2)=\Phi$\\
 3&$\theta^\nu_{12}=\mathcal{O}(\epsilon^2)$&$\varphi_1+\delta^\nu=\Phi$&$2\Phi=-\delta^\nu-\varphi_2$&$\Phi=-\delta^\nu-\varphi_2$\\
4&$\theta^\nu_{23}=\mathcal{O}(\epsilon^3)$&$ \theta^\nu_{23}=\epsilon$& $\theta^\nu_{13}=\epsilon^2$&$\theta^\nu_{13}=\epsilon$\\
5 && &$\theta^\nu_{23}=O(\epsilon^3)$&$\theta^\nu_{23}=\epsilon^2$ (for all $\Phi$) \\
 & & & & or $\theta^\nu_{23}=\mathcal{O}(\epsilon^3)$ \\
 & & & & (for $\Phi$ in $\{\frac{\pi}{2},\frac{3 \pi}{2} \}$)\\
\hline
  \end{tabular}
\end{center}
\mycaption{\label{tab:constraints} Complete set of additional constraints obtained from the textures in \Tab~\ref{tab:textures}. The first column refers to the condition number, the other columns to the different textures.
For texture~2b and $\theta_{23}^\nu=\epsilon^2$, another non-trivial relationship $\varphi_2=f(\Phi)$ can be derived, which can be simplified as $\varphi_2 \simeq -2 \Phi$.}
\end{table}

Our procedure for this section is as follows:
\begin{enumerate}
\item
Given the textures in \Tab~\ref{tab:textures}, we compute the additional constraints on the unphysical parameters. We
require that the textures satisfy \equ{eqlc} and the small angles be only from the set $\{\epsilon,\epsilon^2,0\}$.
We list these additional constraints in \Tab~\ref{tab:constraints}, where the detailed procedure and a discussion of stability can be found in \App~\ref{app:texture}.
\item
 We apply the additional constraints in \Tab~\ref{tab:constraints} to the analytical formulas obtained in \Sec~\ref{sec:obs}. Now the observables depend on the configuration {\em and} texture, $\epsilon$, and a number of unphysical parameters.  Note that we only allow for possibilities within the current experimental bounds. 
\item
 Neglecting higher order corrections, we use the analytical expressions to identify particularly simple relationships among the observables which satisfy our criteria for sum rules in \Sec~\ref{sec:sumrules}. In addition, we identify the allowed regions as a function of the phases $\Phi$ and $\delta^\ell$ for configuration~2, and we relate these phases to (observable) CP violation or conservation.
\end{enumerate}
Note that one should regard the textures in \Tab~\ref{tab:textures}
as the fundamental assumptions for this section, not the mixing angles from the above set, as it should
be clear from \App~\ref{app:texture} (see especially discussion in the very last paragraph).

\begin{table}[tp]
\begin{center}
\begin{tabular}{lllll}
 \hline
 & \multicolumn{2}{c}{{\bf Texture~1a}} & \multicolumn{2}{c}{{\bf Texture~1b}}  \\
Obs. &  Sum rule/Range & (type) & Sum rule/Range & (type)  \\
\hline
$\sin^2 \theta_{13}$ & $\sin^2 \theta_{13} \in [0.02,0.04]$ & &  $ \sin^2 \theta_{13} \simeq 0.04$ \\[0.1cm]
$\sin^2 \theta_{12}$ & $\sin^2 \theta_{12}\simeq 2/(5+2 \sqrt{2})$ & (I) & - &  \\[0.1cm]
& $\sin^2 \theta_{12} \simeq 0.26$ & & Any $\sin^2 \theta_{12}$ in allowed range \\[0.1cm]
$\sin^2 \theta_{23}$ & $\sin^2 \theta_{23} \simeq 2/(5+2 \sqrt{2})$ & (I) & - & \\[0.1cm]
& $\sin^2 \theta_{23} \simeq 0.26$ & &  $\sin^2 \theta_{23} \in [0.34,0.35]$  \\[0.1cm]
$\deltacp$ & $\deltacp \simeq (1-3\sqrt{2}) \delta^\ell$ & (IIIa) & - & \\[0.1cm]
& $\deltacp \in [0,0.6], \, [5.7,2 \pi)$ & & $\deltacp \in [0,0.8], \, [5.5, 2 \pi)$ \\[0.1cm]
$2 \phi_2$ &   $2 \phi_2 \simeq \Phi$ & (IIIb) & $2 \phi_2 \simeq \Phi$ & (IIIb) \\[0.1cm]
 & & & $2 \phi_2 \in [2.5 , 2.7] , \, [3.6,3.8]$ \\[0.1cm]
\hline
\end{tabular}
\end{center}
\mycaption{\label{tab:config1t}
Type~I and~III sum rules according to our classification for textures~1a and~1b, and approximate values/allowed ranges for the observables. The allowed ranges are obtained  for $\epsilon=0.2\simeq \theta_C$, $\Phi$ varied within the allowed range, and all observables within their currently allowed $3\sigma$ ranges (except for texture~1a, where $\sin^2 \theta_{23}$ is slightly below this range).
If no sum rule is given, we have not found one according to our definition. }
\end{table}

For {\bf textures~1a and~1b} from \Tab~\ref{tab:textures}, we find that texture~1a is much more restrictive for
the unphysical parameters than texture~1b (\cf, \Tab~\ref{tab:constraints}). 
For example, we obtain $(\sin^2\theta_{23})_{1a} \simeq 2/(5+2\sqrt{2}) \simeq 0.26$ to zero{\em th} order, which is somewhat below the currently allowed $3 \sigma$ range. Therefore, the texture will be under tension even for the currently allowed $\theta_{23}$ range. In addition, we obtain a lower limit $\sin^2\theta_{13}\gtrsim 0.02$ for texture~1a in comparison to \Sec~\ref{sec:configuration1}, which is within the reach of upcoming reactor and superbeam experiments. 
For texture~1b, however, the observable mixing angles depend on the control parameter $\Phi$. Since, for example,  $\sin^2 \theta_{13}$ and $\sin^2 \theta_{23}$ strongly depend on $\Phi$, it follows that not all possible $\Phi$ are allowed from the current bounds, \ie, we need to approximately have $\Phi$ close to $2.6$ or $3.7$.
For both textures~1a and~1b,  we obtain {\bf type IIIb sum rule} for the Majorana phase $\phi_2$ as
\begin{equation}
 (2 \phi_2)_{1a, \, 1b} \simeq \Phi
\label{equ:sumphi3b}
\end{equation}
 to zero{\it th} order, where
maximal CP violation is obtained for $\Phi=\pi/2$ or $3\pi/2$. On the other hand, $\phi_1$ will not be constrained in any of our examples. One can easily see that for the two textures here by re-writing \equ{Maj} as
\begin{equation}
\label{equ:alpha1}
M_\nu^\text{Maj}=
\text{diag}\left(1,e^{\text{i}\varphi_{1}},e^{\text{i}\varphi_{2}}\right)\cdot\widehat{U}_\nu\cdot\text{diag}\left(\epsilon^2\,e^{\text{2i} \alpha_{1}},\epsilon\,e^{\text{2i}\alpha_{2}}, 1 \right)\cdot\widehat{U}^T_\nu\cdot\text{diag}\left(1,e^{\text{i}\varphi_{1}},e^{\text{i}\varphi_{2}}\right) \, .
\end{equation}
Since $\alpha_1$ always appears as a product with $\epsilon^2$ for the normal hierarchy, and since the textures~1a and~1b do not have entries $\eta^2$, we do not obtain a constraint on $\alpha_1$. From \equ{pmnspara}, we read off that $\alpha_1$ adds to $\phi_1$, which means that $\phi_1$ remains unconstrained.
We summarize the allowed ranges (for any allowed $\Phi$) and sum rules for textures~1a and~1b in \Tab~\ref{tab:config1t}.

For {\bf textures~2a and~2b} from \Tab~\ref{tab:textures}, we know from the discussion in \Sec~\ref{sec:configuration2} that $\sin^2 \theta_{13}$ is zero if all of the small mixing angles 
are chosen to be zero, which means that $\deltacp$ can only be calculated for specific assumptions for the small mixing angles. In addition, $\sin^2\theta_{12}$ tends to be too large (close to maximal) for this configuration.  Therefore, we focus on $\sin^2 \theta_{12}$ first. We obtain for both textures 
\begin{equation}
 (\sin^2\theta_{12})_{2a,2b}=\frac{1}{2}-\frac{1}{\sqrt{2}}\cos{\delta^\ell}\ \theta^\ell_{13}-\frac{1}{\sqrt{2}}\ \theta^\ell_{12}+\mathcal{O}(\epsilon^3) \, .
\label{equ:theta12co}
\end{equation}
We read off from this equation that at least one of the two angles $\theta^\ell_{12}$ and $\theta^\ell_{13}$ has to be of size $\epsilon$, since otherwise $\sin^2\theta_{12}$ violates its currently allowed range (\cf, \equ{constraints}). 
The choices for $\theta_{12}^\ell$ and $\theta_{13}^\ell$ can only be $(\theta_{12}^\ell,\theta_{13}^\ell) \in \{(\epsilon,0), (\epsilon,\epsilon^2),(0,\epsilon), (\epsilon^2,\epsilon), (\epsilon, \epsilon) \}$ from our allowed set of mixing angles. There are then only two small mixing angles remaining: $\theta_{13}^\nu$ and $\theta_{23}^\nu$.
For $\theta_{13}^\nu$, we now have  $\theta_{13}^\nu=\epsilon^2$ (2a) or $\epsilon$ (2b) from \Tab~\ref{tab:constraints}. Therefore, we cannot use $\theta_{13}^\nu=0$, as we have done in the special cases we have studied in \Sec~\ref{sec:configuration2}. 
%
\begin{figure}[p]
\begin{center}
\includegraphics[width=0.7\textwidth]{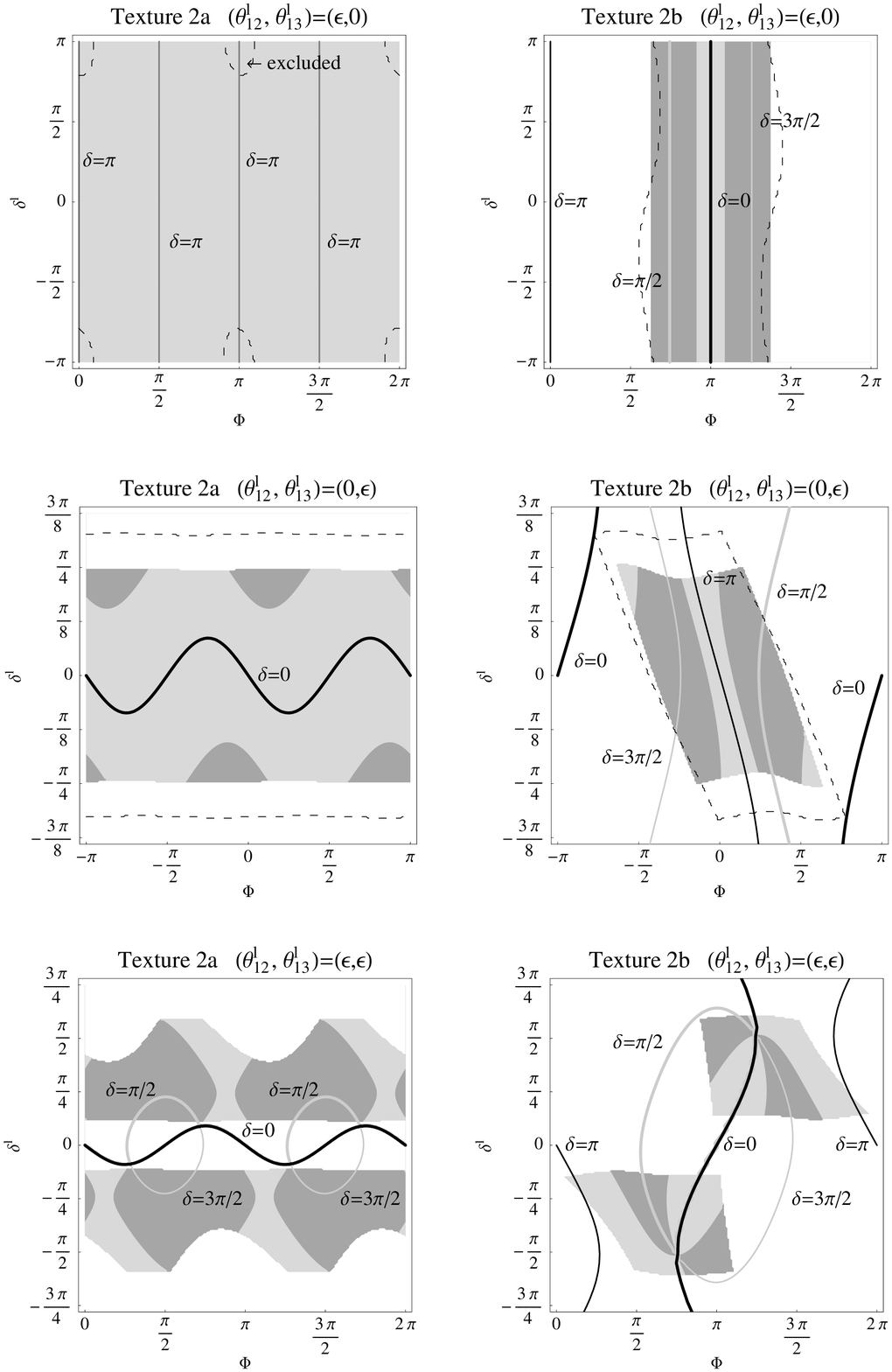}
\end{center}
\mycaption{\label{fig:texture2}
Allowed (shaded) regions in the $\Phi$-$\delta_\ell$ plane for texture~2a (left column) and texture~2b (right column) at the $3\sigma$ confidence level.
The different rows correspond to the different $(\theta_{12}^\ell, \theta_{13}^\ell)$ cases, as given in the panel captions. In the first two rows, the corresponding allowed regions for the perturbed cases $(\theta_{12}^\ell, \theta_{13}^\ell)=(\epsilon,\epsilon^2)$ and $(\epsilon^2,\epsilon)$, respectively, are illustrated as dashed curves.
In each panel, the solid curves mark four different cases for $\deltacp$: $\deltacp=0$ (thick black), $\deltacp=\pi/2$ (thick gray), $\deltacp=\pi$ (thin black), and $\deltacp=3 \pi/2$ (thin gray), \ie, gray curves correspond to maximal (Dirac) CP violation. In addition, the light (dark) shading within the regions refers to $\deltacp$ closer to CP conservation (violation).
}
\end{figure}
 From \equ{theta12co}, we already observe that not every value for $\delta_\ell$ can be chosen if $\theta_{13}^\ell$ is used to obtain a $\sin^2 \theta_{12}$ within the currently allowed range. As we find from the analytical expressions, $\Phi$ plays a similar role especially for $\sin^2 \theta_{13}$. Therefore, we obtain non-trivial constraints in the $\Phi$-$\delta_\ell$ plane from the currently allowed ranges, which will depend on the choices for $\theta_{12}^\ell$ and $\theta_{13}^\ell$.
We show these allowed regions in the $\Phi$-$\delta_\ell$ plane in \figu{texture2} for texture~2a (left column) and texture~2b (right column) as shaded areas.
The different rows correspond to the different $(\theta_{12}^\ell, \theta_{13}^\ell)$ cases, as given in the panel captions. The dashed curves in the upper two rows represent the perturbed cases $(\theta_{12}^\ell,\theta_{13}^\ell)= (\epsilon,\epsilon^2)$ and $(\epsilon^2,\epsilon)$, respectively. Obviously, the deviations from the 
main cases $(\epsilon,0)$ and $(0,\epsilon)$ are small.
 Therefore, we focus on the three cases  $(\theta_{12}^\ell,\theta_{13}^\ell) \in \{(\epsilon,0), (0,\epsilon), (\epsilon, \epsilon) \}$. 

\begin{table}[tp]
\begin{center}
\begin{tabular}{lllll}
 \hline
 & \multicolumn{2}{c}{{\bf Texture~2a}} & \multicolumn{2}{c}{{\bf Texture~2b}}  \\
Obs. &  Sum rule/Range & (type) & Sum rule/Range & (type)  \\
\hline
\multicolumn{5}{l}{$\boldsymbol{(\theta_{12}^\ell, \theta_{13}^\ell) = (\epsilon,0)}$} \\[0.1cm]
$\sin^2 \theta_{13}$ & $\theta_{13}^2 \simeq \frac{1}{2}\ \theta_C^2+\sqrt{2}\cos{(2\Phi)}\ \theta_C^3$ & (IIIb) &  $\theta_{13}^2 \simeq \theta_C^2 \, \left(\frac{3}{2}+\sqrt{2}\cos{\Phi}\right)$ & (IIIb) \\[0.1cm]
& $\sin^2 \theta_{13} \in [0.01,0.03]$ & &  $ \sin^2 \theta_{13} \in[0.003,0.04]$ \\[0.1cm]
$\sin^2 \theta_{12}$ & $\theta_{12}+\frac{1}{\sqrt{2}} \, \theta_C\simeq\frac{\pi}{4}$ & (II) & $\theta_{12}+\frac{1}{\sqrt{2}} \, \theta_C\simeq\frac{\pi}{4}$ & (II) \\[0.1cm]
& $\sin^2 \theta_{12} \simeq 0.36$ & & $\sin^2 \theta_{12} \simeq 0.36$ \\[0.1cm]
$\sin^2 \theta_{23}$ & $\theta_{23}+\frac{1}{4}\,\theta_C^2 \simeq \frac{\pi}{4}$ & (II) &- & \\[0.1cm]
& $\sin^2 \theta_{23} \simeq 0.49$ & &  $\sin^2 \theta_{23} \in [0.48,0.51]$  \\[0.1cm]
$\deltacp$ & $\sin \deltacp \simeq \sqrt{2} \, \theta_C \, \sin 2 \Phi$ & (IIIb) & $\sin \deltacp \simeq \frac{\sqrt{2} \sin \Phi}{\sqrt{3 + 2 \sqrt{2} \cos \Phi}}$ & (IIIb) \\[0.1cm]
& $\deltacp \in [2.9,3.4]$ & & $\deltacp \in [0,1.9]$, $[4.4,2 \pi)$ \\[0.1cm]
$2 \phi_2$ &  $2 \phi_2 \simeq \Phi$ & (IIIb) & $2 \phi_2 \simeq \Phi$ & (IIIb) \\[0.1cm]
& No restriction  & & $2 \phi_2 \in [1.9,4.4]$ & \\[0.1cm]
\hline
\multicolumn{5}{l}{$\boldsymbol{(\theta_{12}^\ell, \theta_{13}^\ell) = (0,\epsilon)}$} \\[0.1cm]
$\sin^2 \theta_{13}$ & $\sin^2 \theta_{13} \in [0.01,0.03]$ & & $\sin^2 \theta_{13} \in [0.003,0.04]$\\[0.1cm]
$\sin^2 \theta_{12}$ & $\theta_{12} + \frac{1}{\sqrt{2}} \, \cos \delta^\ell \, \theta_C \simeq \frac{\pi}{4}$ & (IIIa) & $\theta_{12} + \frac{1}{\sqrt{2}} \, \cos \delta^\ell \, \theta_C \simeq \frac{\pi}{4}$ & (IIIa)  \\[0.1cm]
& $\sin^2 \theta_{12} \in [0.36,0.4]$ &  & $\sin^2 \theta_{12} \in [0.36,0.4]$ & \\[0.1cm]
$\sin^2 \theta_{23}$ & $\theta_{23} - \frac{1}{4} \theta_C^2 \simeq \frac{\pi}{4}$ & (II) & - & \\[0.1cm]
& $\sin^2 \theta_{23} \simeq 0.51$ & & $\sin^2 \theta_{23} \in [0.44,0.54]$  & \\[0.1cm]
$\deltacp$ & $\deltacp \in [-1.1,1.1]$ & & $\deltacp \in [0.4, 2 \pi - 0.4]$ & \\[0.1cm] 
$2 \phi_2$ & $2 \phi_2 \simeq \Phi$ & (IIIb) & $2 \phi_2 \simeq \Phi$ & (IIIb) \\[0.1cm]
& No restriction  & & $2 \phi_2 \in [-2.0,2.0]$ & \\[0.1cm]
\hline
\multicolumn{5}{l}{$\boldsymbol{(\theta_{12}^\ell, \theta_{13}^\ell) = (\epsilon,\epsilon)}$} \\[0.1cm]
$\sin^2 \theta_{12}$ & $\theta_{12} + \frac{1}{\sqrt{2}} \, \theta_C (1+ \cos \delta^\ell) \simeq \frac{\pi}{4}$ & (IIIa) & $\theta_{12} + \frac{1}{\sqrt{2}} \, \theta_C (1+ \cos \delta^\ell) \simeq \frac{\pi}{4}$ & (IIIa)\\[0.1cm]
$\sin^2 \theta_{23}$ & $\theta_{23} - \frac{1}{2}  \theta_C^2 \cos \delta^\ell \simeq \frac{\pi}{4}$ & (IIIa) & - & \\[0.1cm]
& $\sin^2 \theta_{23} \in [0.49,0.52]$ & & $\sin^2 \theta_{23} \in [0.43,0.55]$  & \\[0.1cm]
$\deltacp$ & $\deltacp \in [0.9,2.7], [3.6,5.4]$ & & No restriction &  \\[0.1cm]
$2 \phi_2$ & $2 \phi_2 \simeq \Phi$ & (IIIb) & $2 \phi_2 \simeq \Phi$ & (IIIb) \\[0.1cm]
& No restriction & & $2 \phi_2 \in [0.3,6.0]$ \\[0.1cm]
\hline
\end{tabular}
\end{center}
\mycaption{\label{tab:config2t}
Type~II and~III sum rules according to our classification for textures~2a and~2b, and approximate values/allowed ranges for the observables. The allowed ranges are obtained for $\epsilon=0.2\simeq \theta_C$, $\Phi$ varied within the allowed range, and all observables within their currently allowed $3\sigma$ ranges. The corresponding allowed regions in the $\Phi$-$\delta^\ell$ plane can be found in \figu{texture2}. If no sum rule is given, we have not found one according to our definition.}
\end{table}

Following the procedure outlined above, we find a number of  type~II and~III sum rules and restrictions for the allowed observable ranges for the different $(\theta_{12}^\ell,\theta_{13}^\ell)$ cases. We summarize our results in \Tab~\ref{tab:config2t}. As we have already mentioned in \Sec~\ref{sec:method}, the purpose of the sum rules can be very different, depending on the type. Here the type~II sum rules can be used to falsify a model, whereas the type~III sum rules can be used to actually measure an unphysical or model parameter within the respective model. From \Tab~\ref{tab:config2t} we read off that, depending on the choices for the small mixings angles, either $\Phi$ or $\delta^\ell$ can be reconstructed in many cases~--~such as from a measurement of $\theta_{12}$ or $\theta_{13}$.
Note that the allowed ranges in this table assume that any allowed value can be chosen for $\Phi$. In a specific model, however, $\Phi$ can be a fixed parameter. If the observable is a function of $\Phi$ (and possibly other parameters), the allowed range will be constrained for a specific choice for $\Phi$. In particular, we find that for both textures~2a and~2b, the parameters $\Phi$ and $\delta^\ell$ control the value of $\deltacp$ in all cases. 
We illustrate this dependence for the different cases in \figu{texture2}, where the different curves correspond to $\deltacp=0$ (thick black), $\deltacp=\pi/2$ (thick gray), $\deltacp=\pi$ (thin black), and $\deltacp=3 \pi/2$ (thin gray), \ie, gray curves correspond to maximal (Dirac) CP violation. In addition, the light (dark) shading within the regions refers to $\deltacp$ closer to CP conservation (violation).
In order to construct a model with (Dirac) CP violation, it turns out that texture~2a together with the choice $(\theta_{12}^\ell,\theta_{13}^\ell)=(\epsilon,\epsilon)$ (lower left panel) produces CP violation for any value of $\Phi$. For specific values of $\Phi$, texture~2b works in all cases of $(\theta_{12}^\ell,\theta_{13}^\ell)$ (right column). In these cases, the value of $\deltacp$ will be mainly controlled by $\Phi$, and the value of $\delta^\ell$ is of secondary interest.
The appearance of precisely the unphysical quantities $\Phi$ and $\delta^\ell$ is not accidental. In \Sec~\ref{sec:matrix}, we have assumed that the Majorana neutrino mass matrix has a certain structure related to a model parameter $\Phi$. However, we have not made any similar assumptions for the charged lepton mass matrix. Therefore, it is not surprising that one phase degree of freedom from each the neutrino and charged lepton sectors is remaining. The neutrino sector unphysical phases are, however, strongly determined by $\Phi$ and the texture. If one used similar assumptions for the charged lepton sector, one might be able to constrain $\delta^\ell$ as well.

\section{Summary and conclusions}
\label{sec:summary}

In this study, we have focused on three key aspects: Entangled maximal
mixings in $U_{\mathrm{PMNS}}= U_\ell^\dagger U_\nu$, \ie, maximal
mixing angles in {\em both} $U_\ell$ and $U_\nu$, CP violation in 
complex mass matrices, and a connection to the quark sector. Large mixing 
angles in both $U_\ell$ and $U_\nu$
can be motivated by Froggatt-Nielsen-like models, in which they naturally
appear in both the charged lepton and neutrino sectors. The discussion of 
complex mass matrices can be motivated by the construction 
of CP violating models, and the connection to the quark sector can be
motivated by quark-lepton unification. 

In the first part of the study, we have focused on the aspect of entangled
maximal mixings, and we
have discussed two specific configurations
with mixing angles in $U_\ell$ and $U_\nu$ which can be either maximal, such as
from a symmetry, or small (referred to by stars):
\begin{description}
\item[Configuration~1] with three max. mixing angles $(\theta_{12}^\ell, \theta_{13}^\ell, \theta_{23}^\ell, \theta_{12}^\nu, \theta_{13}^\nu, \theta_{23}^\nu) = (*,\frac{\pi}{4}, \frac{\pi}{4}, *, \frac{\pi}{4}, *)$
\item[Configuration~2] with two max. mixing angles $(\theta_{12}^\ell, \theta_{13}^\ell, \theta_{23}^\ell, \theta_{12}^\nu, \theta_{13}^\nu, \theta_{23}^\nu)=(*,*, \frac{\pi}{4}, \frac{\pi}{4}, *, *)$
\end{description}
We have not made any further specific assumptions for the small mixing angles,
expect that they are as small as they can be used for expansions. 
We have then computed the observables using invariants, and we have discussed
the constraints from their currently allowed ranges.
We have demonstrated that configuration~1 clearly prefers $\sin^2 \theta_{23}$ off maximal mixing at the lower end of the currently allowed range, $\sin^2 \theta_{12}$ close to its best-fit value, $\sin^2 \theta_{13}$ large, and $\deltacp \simeq 0$. Configuration~2, on the other hand, prefers small $\sin^2 \theta_{13}$, $\sin^2 \theta_{12}$ at the upper end of the currently allowed range, and $\sin^2 \theta_{23}$ close to maximal mixing. We have distinguished three different cases for configuration~2, which lead to different results for $\sin^2 \theta_{13}$, $\deltacp$, and $\sin^2 \theta_{23}$. For example, we have found that testing the $\theta_{23}$ octant and $\deltacp$ quadrant sum rules
\begin{equation}
 \theta_{23} \pm \frac{1}{2} \theta_{13}^2 \simeq \frac{\pi}{4} \, ,  \, \qquad \theta_{12} \mp \cos \deltacp \, \theta_{13} \simeq \frac{\pi}{4} 
\label{equ:sum1}
\end{equation}
may reveal if $\theta_{12}^\ell$ (upper signs) or $\theta_{13}^\ell$ (lower signs) is dominating. 
In addition, we have demonstrated how the tri-bimaximal mixing case can be constructed using configuration~2, and how CP violation can be obtained for $\theta_{12}^\ell=\theta_{13}^\ell = (\theta_{23}^\nu)^2$. Our estimate for the required experimental performance to find CP violation in that case requires a CP fraction of 75\%, \ie, CP violation needs to be discovered for 75\% of all (true) values of $\deltacp$.
In addition, the different qualitative cases discussed for configuration~2 have clearly demonstrated that $\stheta$ alone is not sufficient as a performance indicator. If $\sin^2\theta_{13} \gtrsim 0.01$ within the Double Chooz accessible range, $\deltacp$ can separate all the different cases. If $\sin^2 \theta_{13} \lesssim 0.01$, two cases can be excluded. Similarly, a precision $\sin^2 \theta_{23}$ octant measurement has turned out to be a good indicator. 

In the second part of the study, we have linked the aspects
of entangled maximal mixings  and CP violation in complex mass matrices.
We have assumed a normal hierarchy and specific complex neutrino mass textures for the
effective $3\times3$ case, which are of the form
$T_{ij} = \eta^{n_{ij}}$. Here $\eta$ is a complex number and
$n_{ij}$ are real integer numbers.
In this case, a unique phase $\Phi = \arg \eta$ controls the CP violation
coming from the neutrino sector. In a Froggatt-Nielsen-like mechanism, it 
might be introduced as the relative phase between the VEVs of two Standard Model singlet scalar 
flavons fields, if the order one coefficients are assumed
to be real and one field dominates the neutrino mass matrix. In that case, 
the $n_{ij}$'s  are then solely determined by the
quantum numbers of the fermions under a flavor symmetry.
In order to establish a connection to the quark sector
as well, we have, in the (extended) quark-lepton complementarity fashion, assumed that 
$| \eta| = \epsilon \simeq \theta_C$, \ie, $\eta \simeq \theta_C \, \exp(i \Phi)$,
and that all small angles can only be from the sequence $(\epsilon,\epsilon^2, \hdots, 0)$.
An example for such a complex texture can be found in \equ{texture}.

We have demonstrated that the additional assumption of specific mass textures
reduces the parameter space for the unphysical parameters significantly. 
In addition, a number of observables can be related to the model
parameter $\Phi$. For example, the Majorana phase $2\phi_2$ (\cf, definition in \equ{standard}) has turned out to be
approximately $\Phi$ in all cases to leading order, which means that maximal
Majorana CP violation is obtained for $\Phi \in \{\pi/2, 3 \pi/2 \}$.
 For configuration~2, we have again
distinguished different cases, where $\Phi$ and $\delta^\ell$ have turned out
to be the control parameters. We have demonstrated how to construct models with
large CP violation, where, to a first approximation, $\Phi$ controls the specific value of $\deltacp$. However,
in one case, CP conservation can be excluded for any $\Phi$.
We have summarized the allowed ranges in the $\Phi$-$\delta^\ell$ plane in \figu{texture2},
where one can also read off the values of $\deltacp$ as a function of these control parameters.  In addition,
we have summarized all important relationships and observable constraints 
in \Tab~\ref{tab:config2t}.

As a very interesting feature of this study, a number of qualitatively 
different sum rules have emerged. Since the term ``sum rule'' is used in different
contexts in the literature, we have defined it for our study, and we have
classified the sum rules into three categories: Type~I sum rules relate lepton sector
(or quark sector) observables only. They can be used to falsify a model if all of the contributing
observables (at least two) are measured (see, \eg, \equ{sum1}). Type~II sum rules (QLC-type sum rules)
relate at least one lepton sector observable with quark sector observables, such as
the Cabibbo angle. They can be used to falsify a model if all of the contributing 
observables (at least one from the lepton sector) are measured (see, \eg, \Tab~\ref{tab:config2t}). Type~III sum rules relate lepton or quark sector observables with unphysical quantities (type~IIIa), such as the phase $\delta^\ell$, or model parameters (type~IIIb), such as the phase $\Phi$.
They can be used to obtain information on the unphysical quantities or model parameters with a specific model  (see, \eg, \Tab~\ref{tab:config2t}).
In order to qualify as a sum rule, there can be only one such unphysical quantity according to our definition.

In conclusion, we have combined a number of aspects in this study: The concept of entangled
maximal mixings, the concept of complex textures of the form $T_{ij} = \eta^{n_{ij}}$
as powers of a single complex number $\eta$, and the concept of quark-lepton complementarity for
a connection to the quark sector. 
In addition, we have encountered a number of different classes of sum rules,
which we have systematically studied. 
Our concepts have used specific configurations of maximal and
small mixing angles, specific textures, and a normal neutrino mass hierarchy. 
However, they can be applied to different cases in the same way. 
In addition, note that we have considered the full complex case without any {\em a priori} assumptions on the unphysical phases.
Finally, our concept of complex textures allows for a direct connection to discrete flavor symmetry models, such as in \Ref~\cite{Plentinger:2008up} on the one hand side, and for specific testable conclusions for future experiments on the other side. 
These conclusions are not only limited to performance indicators such as $\theta_{13}$ and deviations from maximal mixings, but are also applied to the Dirac and Majorana CP phases. For example, it has turned out that the value of $\deltacp$ is a good performance indicator, even if no CP violation can be established.

\subsubsection*{Acknowledgments}

WW would like to acknowledge support from Emmy Noether program of
Deutsche Forschungsgemeinschaft.

\begin{appendix}

\section{All possible found configurations}
\label{app:allconfigs}

\begin{table}[p]
\begin{center}
\small{
\begin{tabular}{clll}
\hline
\# & $(\theta_{12}^\ell, \theta_{13}^\ell, \theta_{23}^\ell,\theta_{12}^\nu, \theta_{13}^\nu, \theta_{23}^\nu)$
&  $(\theta_{12}^\ell, \theta_{13}^\ell, \theta_{23}^\ell, \delta^\ell, \theta_{12}^\nu, \theta_{13}^\nu, \theta_{23}^\nu,\delta^\nu, \varphi_1, \varphi_2)$ & $(s_{12}^2, s_{13}^2, s_{23}^2, \deltacp)$ \\
\hline
 1 & $(*,*,*,\frac{\pi }{4},*,\frac{\pi }{4})$ & $(\epsilon ,\epsilon ,\epsilon^2,5.30,\frac{\pi }{4},\epsilon ,\frac{\pi }{4},4.71,5.50,1.37)$ &
   $(0.30,0.04,0.50,4.70)$ \\
 2 & $(*,*,\frac{\pi }{4},\frac{\pi }{4},*,*)$ & $(\epsilon ,\epsilon ,\frac{\pi
   }{4},0.59,\frac{\pi }{4},\epsilon ^2,\epsilon ^2,4.91,5.89,0.98)$ &
   $(0.30,0.01,0.50,2.56)$ \\
 3 & $(\frac{\pi }{4},*,\frac{\pi }{4},*,\frac{\pi }{4},*)$ & $(\frac{\pi
   }{4},\epsilon ,\frac{\pi }{4},2.36,\epsilon ^2,\frac{\pi }{4},\epsilon
   ,0.79,1.96,2.16)$ & $(0.30,0.02,0.50,2.93)$ \\
 4 & $(\frac{\pi }{4},*,*,*,\frac{\pi }{4},\frac{\pi }{4})$ & $(\frac{\pi
   }{4},\epsilon ,\epsilon ,0.00,\epsilon ^2,\frac{\pi }{4},\frac{\pi
   }{4},2.75,3.73,4.12)$ & $(0.30,0.02,0.50,3.47)$ \\
 5 & $(*,\frac{\pi }{4},*,*,\frac{\pi }{4},\frac{\pi }{4})$ & $(\epsilon
   ,\frac{\pi }{4},\epsilon ,2.16,\epsilon ^2,\frac{\pi }{4},\frac{\pi
   }{4},5.11,0.59,3.93)$ & $(0.30,0.04,0.43,0.98)$ \\
 6 & $(*,\frac{\pi }{4},\frac{\pi }{4},*,\frac{\pi }{4},*)$ & $(\epsilon
   ,\frac{\pi }{4},\frac{\pi }{4},1.96,\epsilon ^2,\frac{\pi }{4},\epsilon
   ,4.71,0.79,4.12)$ & $(0.30,0.04,0.43,0.95)$ \\
 7 & $(*,*,\frac{\pi }{4},\frac{\pi }{4},*,\frac{\pi }{4})$ & $(\epsilon
   ,\epsilon ,\frac{\pi }{4},0.98,\frac{\pi }{4},\epsilon ,\frac{\pi
   }{4},0.98,6.09,4.52)$ & $(0.30,0.01,0.50,2.87)$ \\
 8 & $(\frac{\pi }{4},\frac{\pi }{4},*,*,\frac{\pi }{4},\frac{\pi }{4})$ &
   $(\frac{\pi }{4},\frac{\pi }{4},\epsilon ,0.00,\epsilon ,\frac{\pi
   }{4},\frac{\pi }{4},0.00,0.39,4.91)$ & $(0.30,0.00,0.50,2.61)$ \\
 9 & $(\frac{\pi }{4},\frac{\pi }{4},\frac{\pi }{4},*,\frac{\pi }{4},*)$ &
   $(\frac{\pi }{4},\frac{\pi }{4},\frac{\pi }{4},4.91,\epsilon ,\frac{\pi
   }{4},\epsilon ^2,0.20,0.20,3.53)$ & $(0.30,0.00,0.50,2.61)$ \\
 10 & $(*,*,*,*,*,\frac{\pi }{4})$ & $(\epsilon ,\epsilon ,\epsilon
   ^2,0.98,\epsilon ,\epsilon ,\frac{\pi }{4},0.59,3.14,0.98)$ &
   $(0.22,0.04,0.49,0.61)$ \\
 11 & $(*,*,\frac{\pi }{4},*,*,*)$ & $(\epsilon ,\epsilon ,\frac{\pi
   }{4},0.00,\epsilon ,\epsilon ,\epsilon ^2,0.79,3.14,3.73)$ &
   $(0.22,0.04,0.50,4.49)$ \\
 12 & $(\frac{\pi }{4},*,*,\frac{\pi }{4},\frac{\pi }{4},\frac{\pi }{4})$ &
   $(\frac{\pi }{4},\epsilon ,\epsilon ,0.79,\frac{\pi }{4},\frac{\pi
   }{4},\frac{\pi }{4},1.18,4.91,5.50)$ & $(0.30,0.04,0.50,0.46)$
   \\
 13 & $(\frac{\pi }{4},*,\frac{\pi }{4},\frac{\pi }{4},\frac{\pi }{4},*)$ &
   $(\frac{\pi }{4},\epsilon ,\frac{\pi }{4},2.36,\frac{\pi }{4},\frac{\pi
   }{4},\epsilon ,1.57,1.18,1.37)$ & $(0.30,0.02,0.50,4.62)$ \\
 14 & $(*,\frac{\pi }{4},\frac{\pi }{4},*,\frac{\pi }{4},\frac{\pi }{4})$ &
   $(\epsilon ,\frac{\pi }{4},\frac{\pi }{4},4.32,\epsilon ^2,\frac{\pi
   }{4},\frac{\pi }{4},1.57,3.93,1.57)$ & $(0.30,0.04,0.50,6.05)$
   \\
 15 & $(\frac{\pi }{4},\frac{\pi }{4},\frac{\pi }{4},\frac{\pi }{4},*,*)$ &
   $(\frac{\pi }{4},\frac{\pi }{4},\frac{\pi }{4},0.39,\frac{\pi }{4},\epsilon
   ,\epsilon ,1.18,5.30,4.91)$ & $(0.30,0.02,0.50,3.88)$ \\
 16 & $(\frac{\pi }{4},\frac{\pi }{4},*,\frac{\pi }{4},*,\frac{\pi }{4})$ &
   $(\frac{\pi }{4},\frac{\pi }{4},\epsilon ,2.95,\frac{\pi }{4},\epsilon
   ,\frac{\pi }{4},3.73,5.30,4.91)$ & $(0.30,0.02,0.50,3.89)$ \\
 17 & $(\frac{\pi }{4},*,\frac{\pi }{4},*,\frac{\pi }{4},\frac{\pi }{4})$ &
   $(\frac{\pi }{4},\epsilon ,\frac{\pi }{4},2.16,\epsilon ^2,\frac{\pi
   }{4},\frac{\pi }{4},5.89,1.37,2.55)$ & $(0.30,0.02,0.50,3.11)$
   \\
 18 & $(*,\frac{\pi }{4},*,\frac{\pi }{4},\frac{\pi }{4},\frac{\pi }{4})$ &
   $(\epsilon ,\frac{\pi }{4},\epsilon ,4.12,\frac{\pi }{4},\frac{\pi
   }{4},\frac{\pi }{4},1.37,5.50,2.16)$ & $(0.31,0.04,0.43,0.67)$
   \\
 19 & $(*,\frac{\pi }{4},\frac{\pi }{4},\frac{\pi }{4},\frac{\pi }{4},*)$ &
   $(\epsilon ,\frac{\pi }{4},\frac{\pi }{4},2.16,\frac{\pi }{4},\frac{\pi
   }{4},\epsilon ,4.71,0.98,4.32)$ & $(0.29,0.04,0.43,5.60)$ \\
 20 & $(\frac{\pi }{4},\frac{\pi }{4},\frac{\pi }{4},*,\frac{\pi }{4},\frac{\pi }{4})$
   & $(\frac{\pi }{4},\frac{\pi }{4},\frac{\pi }{4},2.95,\epsilon ,\frac{\pi
   }{4},\frac{\pi }{4},4.32,0.59,5.11)$ & $(0.30,0.00,0.50,4.66)$ \\
 21 & $(\frac{\pi }{4},\frac{\pi }{4},*,\frac{\pi }{4},\frac{\pi }{4},\frac{\pi }{4})$
   & $(\frac{\pi }{4},\frac{\pi }{4},\epsilon ,1.57,\frac{\pi }{4},\frac{\pi
   }{4},\frac{\pi }{4},3.93,1.96,4.91)$ & $(0.30,0.02,0.50,6.14)$
   \\
 22 & $(\frac{\pi }{4},\frac{\pi }{4},\frac{\pi }{4},\frac{\pi }{4},\frac{\pi }{4},*)$
   & $(\frac{\pi }{4},\frac{\pi }{4},\frac{\pi }{4},5.11,\frac{\pi }{4},\frac{\pi
   }{4},\epsilon ,2.95,0.59,0.79)$ & $(0.30,0.02,0.50,5.28)$ \\
 23 & $(\frac{\pi }{4},\frac{\pi }{4},\frac{\pi }{4},\frac{\pi }{4},*,\frac{\pi }{4})$
   & $(\frac{\pi }{4},\frac{\pi }{4},\frac{\pi }{4},2.16,\frac{\pi }{4},\epsilon
   ,\frac{\pi }{4},3.73,5.89,4.71)$ & $(0.30,0.03,0.50,3.37)$ \\
 24 & $(*,\frac{\pi }{4},*,*,*,\frac{\pi }{4})$ & $(\epsilon ,\frac{\pi
   }{4},\epsilon ^2,4.91,\epsilon ,\epsilon ,\frac{\pi
   }{4},2.36,0.79,2.55)$ & $(0.29,0.04,0.59,3.50)$ \\
 25 & $(*,\frac{\pi }{4},\frac{\pi }{4},*,*,*)$ & $(\epsilon ,\frac{\pi
   }{4},\frac{\pi }{4},6.09,\epsilon ,\epsilon ,\epsilon ^2,5.30,5.30,0.79)$
   & $(0.30,0.04,0.59,2.75)$ \\
 26 & $(\frac{\pi }{4},\frac{\pi }{4},*,\frac{\pi }{4},*,*)$ & $(\frac{\pi
   }{4},\frac{\pi }{4},\epsilon ,5.89,\frac{\pi }{4},\epsilon ,\epsilon
   ,0.98,0.98,3.73)$ & $(0.30,0.03,0.49,4.73)$ \\
 27 & $(\frac{\pi }{4},*,\frac{\pi }{4},*,*,\frac{\pi }{4})$ & $(\frac{\pi
   }{4},\epsilon ,\frac{\pi }{4},4.52,\epsilon ,\epsilon ,\frac{\pi
   }{4},2.55,4.71,6.09)$ & $(0.30,0.04,0.48,6.05)$ \\
 28 & $(*,*,\frac{\pi }{4},*,*,\frac{\pi }{4})$ & $(\epsilon ,\epsilon ,\frac{\pi
   }{4},1.57,\epsilon ,\epsilon ,\frac{\pi }{4},3.53,3.93,2.36)$ &
   $(0.22,0.04,0.50,5.12)$ \\
 29 & $(\frac{\pi }{4},*,*,*,*,\frac{\pi }{4})$ & $(\frac{\pi }{4},\epsilon
   ,\epsilon ,2.16,\epsilon ,\epsilon ,\frac{\pi }{4},5.11,0.98,0.00)$ &
   $(0.30,0.03,0.48,0.29)$ \\
 30 & $(\frac{\pi }{4},*,*,*,*,*)$ & $(\frac{\pi }{4},\epsilon ,\epsilon
   ,0.00,\epsilon ,\epsilon ,\epsilon ,0.00,0.00,3.14)$ &
   $(0.30,0.00,0.29,0.00)$ \\
 31 & $(\frac{\pi }{4},*,\frac{\pi }{4},*,*,*)$ & $(\frac{\pi }{4},\epsilon
   ,\frac{\pi }{4},0.39,\epsilon ,\epsilon ,\epsilon ,3.34,0.98,0.00)$ &
   $(0.30,0.03,0.48,0.27)$ \\
 32 & $(\frac{\pi }{4},*,\frac{\pi }{4},\frac{\pi }{4},\frac{\pi }{4},\frac{\pi }{4})$
   & $(\frac{\pi }{4},\epsilon ,\frac{\pi }{4},5.69,\frac{\pi }{4},\frac{\pi
   }{4},\frac{\pi }{4},0.00,5.69,4.52)$ & $(0.30,0.03,0.50,2.27)$ \\
 33 & $(\frac{\pi }{4},\frac{\pi }{4},\frac{\pi }{4},\frac{\pi }{4},\frac{\pi
   }{4},\frac{\pi }{4})$ & $(\frac{\pi }{4},\frac{\pi }{4},\frac{\pi
   }{4},3.14,\frac{\pi }{4},\frac{\pi }{4},\frac{\pi }{4},2.36,2.75,0.98)$
   & $(0.30,0.01,0.50,5.89)$ \\
 34 & $(*,\frac{\pi }{4},\frac{\pi }{4},\frac{\pi }{4},\frac{\pi }{4},\frac{\pi }{4})$
   & $(\epsilon ,\frac{\pi }{4},\frac{\pi }{4},5.11,\frac{\pi }{4},\frac{\pi
   }{4},\frac{\pi }{4},0.20,5.89,3.53)$ & $(0.30,0.03,0.51,1.53)$
   \\
 35 & $(*,\frac{\pi }{4},*,\frac{\pi }{4},*,\frac{\pi }{4})$ & $(\epsilon
   ,\frac{\pi }{4},\epsilon ^2,1.37,\frac{\pi }{4},\epsilon ,\frac{\pi
   }{4},1.96,1.18,5.69)$ & $(0.29,0.04,0.59,4.76)$ \\
 36 & $(*,\frac{\pi }{4},\frac{\pi }{4},\frac{\pi }{4},*,*)$ & $(\epsilon
   ,\frac{\pi }{4},\frac{\pi }{4},0.00,\frac{\pi }{4},\epsilon ,\epsilon^2,2.95,5.11,3.34)$ 
& $(0.30,0.04,0.59,1.41)$ \\
 37 & $(*,\frac{\pi }{4},\frac{\pi }{4},*,*,\frac{\pi }{4})$ & $(\epsilon
   ,\frac{\pi }{4},\frac{\pi }{4},0.98,\epsilon ,\epsilon ,\frac{\pi
   }{4},0.79,1.18,5.69)$ & $(0.29,0.04,0.65,2.85)$ \\
 38 & $(\frac{\pi }{4},*,*,\frac{\pi }{4},*,\frac{\pi }{4})$ & $(\frac{\pi
   }{4},\epsilon ,\epsilon ,2.16,\frac{\pi }{4},\epsilon ,\frac{\pi
   }{4},0.98,5.30,4.32)$ & $(0.30,0.04,0.48,4.47)$ \\
 39 & $(\frac{\pi }{4},*,\frac{\pi }{4},\frac{\pi }{4},*,*)$ & $(\frac{\pi
   }{4},\epsilon ,\frac{\pi }{4},0.39,\frac{\pi }{4},\epsilon ,\epsilon
   ,5.50,5.30,4.32)$ & $(0.30,0.04,0.48,4.44)$ \\
 40 & $(\frac{\pi }{4},*,*,\frac{\pi }{4},*,*)$ & $(\frac{\pi }{4},\epsilon
   ,\epsilon ,0.00,\frac{\pi }{4},\epsilon ,\epsilon ,1.18,5.11,1.96)$ &
   $(0.31,0.00,0.29,4.71)$ \\
 41 & $(\frac{\pi }{4},*,\frac{\pi }{4},\frac{\pi }{4},*,\frac{\pi }{4})$ &
   $(\frac{\pi }{4},\epsilon ,\frac{\pi }{4},1.37,\frac{\pi }{4},\epsilon
   ,\frac{\pi }{4},3.53,1.77,0.39)$ & $(0.29,0.04,0.49,1.89)$ \\
 42 & $(*,\frac{\pi }{4},\frac{\pi }{4},\frac{\pi }{4},*,\frac{\pi }{4})$ &
   $(\epsilon ,\frac{\pi }{4},\frac{\pi }{4},0.98,\frac{\pi }{4},\epsilon
   ,\frac{\pi }{4},0.20,1.57,6.09)$ & $(0.30,0.03,0.65,4.34)$ \\
 43 & $(\frac{\pi }{4},\frac{\pi }{4},*,\frac{\pi }{4},\frac{\pi }{4},*)$ &
   $(\frac{\pi }{4},\frac{\pi }{4},\epsilon ,0.98,\frac{\pi }{4},\frac{\pi
   }{4},\epsilon ,3.53,1.57,4.91)$ & $(0.30,0.04,0.34,3.77)$ \\
 44 & $(\frac{\pi }{4},\frac{\pi }{4},*,*,\frac{\pi }{4},*)$ & $(\frac{\pi
   }{4},\frac{\pi }{4},\epsilon ,1.57,\epsilon ,\frac{\pi }{4},\epsilon
   ,4.91,0.59,3.93)$ & $(0.31,0.02,0.27,3.12)$ \\
\hline
\end{tabular}
} 
\end{center}
\vspace*{-0.5cm}
\mycaption{\label{tab:allconfigs}All possible configurations (and implementation examples) found in the study. See main text for more explanations.
}
\end{table}

In \Tab~\ref{tab:allconfigs}, we list all possible configurations that we have found by choosing the
mixing angles from the set $\{\pi/4, \epsilon, \epsilon^2, 0 \}$ (for $\epsilon=0.2$). In this table, we list the configuration, a possible implementation with a set of specific unphysical parameters in \equ{params}, and the corresponding physical observables. We require that the implementations pass the selection criterion 
\begin{equation}
S \equiv \left( \frac{\sin^2 \theta_{12}  - 0.3}{0.3 \times 0.09} \right)^2 + \left( \frac{\sin^2 \theta_{23} -  0.5}{0.5 \times 0.16} \right )^2 \le 11.83, \quad \sin^2 \theta_{13}
\le 0.04 \, .
\label{equ:selector}
\end{equation}
This selector corresponds to compatibility at the $3\sigma$ confidence level (\cf, \equ{constraints})
with a Gaussian $\chi^2$  estimate
for $\sin^2 \theta_{12}$ and $\sin^2 \theta_{23}$ (2 d.o.f.), and a hard cut for $\sin^2 \theta_{13}$.
 We do not show $\alpha_1$ and $\alpha_2$, which  can take any value because these phases simply add to the Majorana phases $\phi_1$ and $\phi_2$ for which we have not imposed any constraints yet. For the same reason, we do not show $\phi_1$ and $\phi_2$.
From the table, we can read off that we find a valid implementation for $U_{\mathrm{PMNS}}$ for $44$ out of the theoretically possible $2^6=64$ configurations. Many of these configurations lead, however, to anarchic or semi-anarchic textures.
The particular (subjective) choice of the two configurations in this paper is based on the following criteria:
\begin{itemize}
\item
 We have maximal mixings in both $U_\ell$ and $U_\nu$
\item
 We obtain interesting textures from the configuration, such as
no anarchic patterns\footnote{Specifically, configurations with many maximal mixing angles
tend to produce anarchic textures.}
\item
 We find solutions for the corresponding textures satisfying \equ{eqlc} for specific $\Phi$
\item 
 We find solutions for $M_\ell$ as well satisfying these criteria (which is not a necessary condition,
but might be required depending on the model) 
\item
 We find a strong impact of the condition \equ{eqlc} together with the corresponding textures
on the allowed observable ranges
\end{itemize}
Therefore, we choose configurations \#2 and \#6 from this list for a more detailed analysis in this study.

\section{Additional constraints from textures}
\label{app:texture}

Here we clarify our definition of a texture, and we demonstrate how we derive the additional constraints coming from \equ{eqlc}. First of all, we refer to the effective Majorana
neutrino matrix entries as $(M_\nu^\text{Maj})_{ij}$, whereas the texture entries are referred to as $T_{ij}$.
In order to obtain the texture entries, we expand 
\begin{equation}
 (M_\nu^\text{Maj})_{ij}=M^{(0)}_{ij}+M^{(1)}_{ij} \, \epsilon+M^{(2)}_{ij} \, \epsilon^2 + \mathcal{O}(\epsilon^3) \, ,
\label{equ:mexp}
\end{equation}
with $\epsilon\simeq\theta_C$ real, which means that the $M^{(k)}_{ij}$'s are complex numbers. 
Let the first non-vanishing coefficient for any $i$, $j$ be $M^{(n_{ij})}_{ij}$, \ie, $|M^{(n_{ij})}_{ij}|>0$ and $|M^{(r)}_{ij}|=0$ for all $r<n_{ij}$. Then, if \equ{eqlc} holds, the leading order term  can be written as
\begin{equation} 
M^{(n_{ij})}_{ij} \, \epsilon^{n_{ij}} =  e^{i\Pi} \, M \, K_{ij} \, \eta^{n_{ij}}= e^{i\Pi} \, M \, K_{ij} \, \left( \epsilon^{n_{ij}} \, e^{i\Phi n_{ij}} \right) \equiv  e^{i\Pi} \, M \, K_{ij}  \, T_{ij}
\label{equ:leading}
\end{equation}
with $K_{ij}>0$, real, and order unity, $\Pi$ a global phase independent of $i$ and $j$, and $M$ the absolute neutrino mass scale. Therefore, the texture entry $T_{ij}$ is defined as $T_{ij} \equiv \eta^{n_{ij}}$. The matrix is therefore determined by the texture entries $T_{ij}$ except from a global phase (which can be absorbed in the field definitions), the absolute neutrino mass scale, real order one coefficients $K_{ij}$, and higher order corrections.\footnote{Here the underlying assumption is that the complex phases in $M_\nu^\text{Maj}$, and therefore the observables, are to leading order determined by $\eta^{n_{ij}}$ (\cf, \equ{fn}). However, higher order corrections are used in \Refs~\cite{Plentinger:2006nb,Plentinger:2007px,Winter:2007yi} to fit the data, which means that they need to be absorbed in the leading order one coefficients $K_{ij}$. In the complex case, these higher order corrections can have complex coefficients, which means that the leading order $K_{ij}$'s are, in practice, not real but only ``sufficiently real''. One can show that ``sufficiently real'' means that $|\mathrm{arg}(K_{ij})| \lesssim \theta_C \simeq 12^\circ$. Our conclusions for the phases will therefore only hold up to this precision.} If all $M^{(k)}_{ij}=0$ for $0 \le k \le 2$,
we use a texture ``zero'' with an undefined phase. This cutoff can be motivated by the current measurement precision, which does not allow for the discrimination of higher order terms. Consequently, we only choose values $\epsilon$, $\epsilon^2$, and $0$ for the small mixing angles, where the zero refers to mixing angles $\mathcal{O}(\epsilon^3)$.\footnote{In some cases, we will use expansions up to the third order if there are no zero{\it th} and first order terms. In this case, the contributions of the third order terms could be sizeable.}
 
As the next step, let us illustrate how we obtain additional constraints from the textures. For a given configuration, we compute $(M_{\nu}^\text{Maj})_{ij}$ in \equ{Maj} using $M_\nu^\text{diag} \propto (\epsilon^2,\epsilon,1)$
and $U_\nu$ from the respective configuration. 
We apply \equ{mexp} to $(M_{\nu}^\text{Maj})_{ij}$ and identify the leading order term. Then we obtain the texture entry $T_{ij}$ from \equ{leading} by neglecting $M \, K_{ij}$ (we keep the global phase to explicitely eliminate one of the phases). We refer to this texture extraction mechanism as ``$(M_{\nu}^\text{Maj})_{ij} \rightarrow e^{i \Pi} \, T_{ij}$''.
The texture entry $T_{ij}$ has then to be matched to the corresponding texture in \Tab~\ref{tab:textures}, which leads to additional constraints for the neutrino parameters. These constraints may in some cases depend on $\Phi$.
Note that not all of the neutrino parameters can be constrained by this procedure, and that the charged lepton parameters in $U_\ell$ remain untouched. One could extend this procedure to the charged lepton sector with corresponding textures, but this would be way more complicated because of the additional freedom coming  from $U_\ell'$ in \equ{lepdiag}.
In addition, in Froggatt-Nielsen-like models, a second field could be present in the charged lepton mass matrix in order to produce the stronger hierarchy.

Let us illustrate the texture extraction with several qualitatively different examples from configuration~1 (the examples hold for both textures~1a and~1b). First of all, we compute $(M_{\nu}^\text{Maj})_{ij}$ from \equ{Maj}, and expand it according to \equ{mexp}. For the 11-element, it is sufficient to expand  to zero{\it th} order as
\begin{equation}\label{equ:m11}
(M^\text{Maj}_{\nu})_{11}=\frac{1}{2}e^{-2i\delta^\nu}+\mathcal{O}(\epsilon)\rightarrow e^{i \Pi} \, T_{11}= e^{i \Pi} \, 1 \, ,
\end{equation}
where the ``1'' on the right is the 11-texture entry $T_{11}$ from \Tab~\ref{tab:textures}.
Therefore, we can fix the global phase as $\Pi=-2\delta^\nu$, and we do not obtain any conditions for the angles.
Similar to that, we will always choose a texture entry $T_{ij}=1$ to fix the global phase.
We then expand the element $(M_{\nu}^\text{Maj})_{22}$ to first order as
\begin{equation}\label{equ:m22}
  (M_{\nu}^\text{Maj})_{22}=e^{2i(\alpha_2+\varphi_1)} \, \epsilon +\mathcal{O}(\epsilon^2)\rightarrow e^{i \Pi} \, T_{22}=e^{i \Pi} \, \epsilon \, e^{i\Phi} \, .\\
\end{equation}
Here the $\epsilon$ comes from the diagonal neutrino mass matrix $M_\nu^\text{diag}$. Again, we obtain no condition for the angles, but only for the phases: $2(\alpha_2+\varphi_1)=\Pi+\Phi$, or $2(\alpha_2+\varphi_1+\delta_\nu)=\Phi$. It relates a combination of unphysical phases to the model parameter $\Phi$. 
The next element $(M_{\nu}^\text{Maj})_{12}$ is, to second order, given by
\begin{equation}
\label{equ:exm3}
  (M_{\nu}^\text{Maj})_{12}=\frac{1}{2}\ e^{i(\varphi_1-\delta^\nu)} \, \theta_{23}^\nu+\frac{1}{\sqrt{2}}\ e^{i(2\alpha_2+\varphi_1)} \, \theta_{12}^\nu \, \epsilon +\mathcal{O}(\epsilon^3)\rightarrow 
\begin{cases}
T_{12}=0 & \text{(1a)}\\
e^{i \Pi} \, T_{12}=e^{i\Pi} \, \eta & \text{(1b)}                                                                                                        \end{cases}
\end{equation}
This example nicely demonstrates the appearance of a mixed $\theta_{12}^\nu \, \epsilon$-term, where the $\epsilon$ comes from the neutrino mass hierarchy $M_\nu^\text{diag}=(\epsilon^2,\epsilon,1)$. Compared to the previous section, where we assumed the small mixing angles to be $\mathcal{O}(\epsilon)$ in order to expand in them, we now have a discrete set of possibilities $\theta^\nu_{ij}$, $\theta^\ell_{ij} \in \{\epsilon, \epsilon^2, 0 \}$. In the expansion \equ{mexp}, however, the order of $\theta^\nu_{ij}$ is not specified without additional assumptions. Therefore, the initial expansion will contain $\epsilon$'s as well as $\theta^\nu_{ij}$'s. 
For texture~1a, $T_{12}$ is zero, which means that all terms of the order lower than $\epsilon^3$ have to vanish.
Therefore, we have to choose $\theta^\nu_{23}=\mathcal{O}(\epsilon^3)$ and $\theta^\nu_{12}=\mathcal{O}(\epsilon^2)$ as that they only appear in combinations $\mathcal{O}(\epsilon^3)$.
For texture~1b, $T_{12}$ is $\eta$, which means that $(M_{\nu}^\text{Maj})_{12}$ must have a non-vanishing first order term. That can only come from $\theta^\nu_{23}=\epsilon$ because $\theta^\nu_{12}=\mathcal{O}(\epsilon)$. For the phases, we obtain in this case $\varphi_1-\delta^\nu=\Pi+\Phi$, or $\varphi_1+ \delta^\nu = \Phi$. 

We calculate the conditions on the angles and the phases for the other matrix elements and textures in the same way. The complete list of results are given in \Tab~\ref{tab:constraints}. 
Note that we do not use cancellations of terms to derive these constraints. For example, in \equ{exm3}, one may as well choose $\theta_{23}^\nu=\epsilon$, $\theta_{12}^\nu=\epsilon/\sqrt{2}$, and appropriate phases to cancel the two terms and have $T_{12}=0$.
In this case, $\theta_{12}^\nu$ is not chosen from the set $\{\epsilon,\epsilon^2,0\}$, but is only of the order $\epsilon$. 
However, as a consequence, the conditions derived from this choice would not be stable under the variation of the angles by order one coefficients. For our choices, the angles and the mass hierarchy entries can be varied by order one coefficients, and therefore, our conditions are stable. Conversely, a variation of the order one coefficients in the textures does not change our conclusions as long as the phases of these variations are $\mathcal{O}(\theta_C)$.
This aspect might be relevant for Froggatt-Nielsen-like models, where, in general, arbitrary order one coefficients are allowed. The stability criterion will hold until we use a specific $\epsilon=0.2$ for numerical estimates, or we replace $\epsilon \rightarrow \theta_C$ to derive type~II (or type~III) sum rules.

\end{appendix}


\begin{thebibliography}{10}
\expandafter\ifx\csname bibnamefont\endcsname\relax
  \def\bibnamefont#1{#1}\fi
\expandafter\ifx\csname bibfnamefont\endcsname\relax
  \def\bibfnamefont#1{#1}\fi
\expandafter\ifx\csname url\endcsname\relax
  \def\url#1{\texttt{#1}}\fi
\expandafter\ifx\csname urlprefix\endcsname\relax\def\urlprefix{URL }\fi
\providecommand{\bibinfo}[2]{#2}
\providecommand{\eprint}[2][]{\url{#2}}

\bibitem{Frampton:2004ud}
\bibinfo{author}{\bibfnamefont{P.~H.} \bibnamefont{Frampton}},
  \bibinfo{author}{\bibfnamefont{S.~T.} \bibnamefont{Petcov}},
  \bibnamefont{and}
  \bibinfo{author}{\bibfnamefont{W.}~\bibnamefont{Rodejohann}},
  \bibinfo{journal}{Nucl. Phys.} \textbf{\bibinfo{volume}{B687}},
  \bibinfo{pages}{31} (\bibinfo{year}{2004}), \eprint{hep-ph/0401206}.

\bibitem{Plentinger:2006nb}
\bibinfo{author}{\bibfnamefont{F.}~\bibnamefont{Plentinger}},
  \bibinfo{author}{\bibfnamefont{G.}~\bibnamefont{Seidl}}, \bibnamefont{and}
  \bibinfo{author}{\bibfnamefont{W.}~\bibnamefont{Winter}},
  \bibinfo{journal}{Nucl. Phys.} \textbf{\bibinfo{volume}{B791}},
  \bibinfo{pages}{60} (\bibinfo{year}{2008}), \eprint{hep-ph/0612169}.

\bibitem{Plentinger:2007px}
\bibinfo{author}{\bibfnamefont{F.}~\bibnamefont{Plentinger}},
  \bibinfo{author}{\bibfnamefont{G.}~\bibnamefont{Seidl}}, \bibnamefont{and}
  \bibinfo{author}{\bibfnamefont{W.}~\bibnamefont{Winter}},
  \bibinfo{journal}{Phys. Rev.} \textbf{\bibinfo{volume}{D76}},
  \bibinfo{pages}{113003} (\bibinfo{year}{2007}), \eprint{arXiv:0707.2379
  [hep-ph]}.

\bibitem{Froggatt:1978nt}
\bibinfo{author}{\bibfnamefont{C.~D.} \bibnamefont{Froggatt}} \bibnamefont{and}
  \bibinfo{author}{\bibfnamefont{H.~B.} \bibnamefont{Nielsen}},
  \bibinfo{journal}{Nucl. Phys.} \textbf{\bibinfo{volume}{B147}},
  \bibinfo{pages}{277} (\bibinfo{year}{1979}).

\bibitem{Enkhbat:2005xb}
\bibinfo{author}{\bibfnamefont{T.}~\bibnamefont{Enkhbat}} \bibnamefont{and}
  \bibinfo{author}{\bibfnamefont{G.}~\bibnamefont{Seidl}},
  \bibinfo{journal}{Nucl. Phys.} \textbf{\bibinfo{volume}{B730}},
  \bibinfo{pages}{223} (\bibinfo{year}{2005}), \eprint{hep-ph/0504104}.

\bibitem{Altarelli:2004jb}
\bibinfo{author}{\bibfnamefont{G.}~\bibnamefont{Altarelli}},
  \bibinfo{author}{\bibfnamefont{F.}~\bibnamefont{Feruglio}}, \bibnamefont{and}
  \bibinfo{author}{\bibfnamefont{I.}~\bibnamefont{Masina}},
  \bibinfo{journal}{Nucl. Phys.} \textbf{\bibinfo{volume}{B689}},
  \bibinfo{pages}{157} (\bibinfo{year}{2004}), \eprint{hep-ph/0402155}.

\bibitem{Romanino:2004ww}
\bibinfo{author}{\bibfnamefont{A.}~\bibnamefont{Romanino}},
  \bibinfo{journal}{Phys. Rev.} \textbf{\bibinfo{volume}{D70}},
  \bibinfo{pages}{013003} (\bibinfo{year}{2004}), \eprint{hep-ph/0402258}.

\bibitem{Antusch:2004re}
\bibinfo{author}{\bibfnamefont{S.}~\bibnamefont{Antusch}} \bibnamefont{and}
  \bibinfo{author}{\bibfnamefont{S.~F.} \bibnamefont{King}},
  \bibinfo{journal}{Phys. Lett.} \textbf{\bibinfo{volume}{B591}},
  \bibinfo{pages}{104} (\bibinfo{year}{2004}), \eprint{hep-ph/0403053}.

\bibitem{deS.Pires:2004bq}
\bibinfo{author}{\bibfnamefont{C.~A.} \bibnamefont{de~S.~Pires}},
  \bibinfo{journal}{J. Phys.} \textbf{\bibinfo{volume}{G30}},
  \bibinfo{pages}{B29} (\bibinfo{year}{2004}), \eprint{hep-ph/0404146}.

\bibitem{Li:2005yj}
\bibinfo{author}{\bibfnamefont{N.}~\bibnamefont{Li}} \bibnamefont{and}
  \bibinfo{author}{\bibfnamefont{B.-Q.} \bibnamefont{Ma}},
  \bibinfo{journal}{Eur. Phys. J.} \textbf{\bibinfo{volume}{C42}},
  \bibinfo{pages}{17} (\bibinfo{year}{2005}), \eprint{hep-ph/0504161}.

\bibitem{Ohlsson:2005js}
\bibinfo{author}{\bibfnamefont{T.}~\bibnamefont{Ohlsson}},
  \bibinfo{journal}{Phys. Lett.} \textbf{\bibinfo{volume}{B622}},
  \bibinfo{pages}{159} (\bibinfo{year}{2005}), \eprint{hep-ph/0506094}.

\bibitem{Hochmuth:2007wq}
\bibinfo{author}{\bibfnamefont{K.~A.} \bibnamefont{Hochmuth}},
  \bibinfo{author}{\bibfnamefont{S.~T.} \bibnamefont{Petcov}},
  \bibnamefont{and}
  \bibinfo{author}{\bibfnamefont{W.}~\bibnamefont{Rodejohann}},
  \bibinfo{journal}{Phys. Lett.} \textbf{\bibinfo{volume}{B654}},
  \bibinfo{pages}{177} (\bibinfo{year}{2007}), \eprint{arXiv:0706.2975
  [hep-ph]}.

\bibitem{Masina:2005hf}
\bibinfo{author}{\bibfnamefont{I.}~\bibnamefont{Masina}},
  \bibinfo{journal}{Phys. Lett.} \textbf{\bibinfo{volume}{B633}},
  \bibinfo{pages}{134} (\bibinfo{year}{2006}), \eprint{hep-ph/0508031}.

\bibitem{Petcov:1993rk}
\bibinfo{author}{\bibfnamefont{S.~T.} \bibnamefont{Petcov}} \bibnamefont{and}
  \bibinfo{author}{\bibfnamefont{A.~Y.} \bibnamefont{Smirnov}},
  \bibinfo{journal}{Phys. Lett.} \textbf{\bibinfo{volume}{B322}},
  \bibinfo{pages}{109} (\bibinfo{year}{1994}), \eprint{hep-ph/9311204}.

\bibitem{Smirnov:2004ju}
\bibinfo{author}{\bibfnamefont{A.~Y.} \bibnamefont{Smirnov}}
  (\bibinfo{year}{2004}), \eprint{hep-ph/0402264}.

\bibitem{Raidal:2004iw}
\bibinfo{author}{\bibfnamefont{M.}~\bibnamefont{Raidal}},
  \bibinfo{journal}{Phys. Rev. Lett.} \textbf{\bibinfo{volume}{93}},
  \bibinfo{pages}{161801} (\bibinfo{year}{2004}), \eprint{hep-ph/0404046}.

\bibitem{Minakata:2004xt}
\bibinfo{author}{\bibfnamefont{H.}~\bibnamefont{Minakata}} \bibnamefont{and}
  \bibinfo{author}{\bibfnamefont{A.~Y.} \bibnamefont{Smirnov}},
  \bibinfo{journal}{Phys. Rev.} \textbf{\bibinfo{volume}{D70}},
  \bibinfo{pages}{073009} (\bibinfo{year}{2004}), \eprint{hep-ph/0405088}.

\bibitem{Jezabek:1999ta}
\bibinfo{author}{\bibfnamefont{M.}~\bibnamefont{Jezabek}} \bibnamefont{and}
  \bibinfo{author}{\bibfnamefont{Y.}~\bibnamefont{Sumino}},
  \bibinfo{journal}{Phys. Lett.} \textbf{\bibinfo{volume}{B457}},
  \bibinfo{pages}{139} (\bibinfo{year}{1999}), \eprint{hep-ph/9904382}.

\bibitem{Giunti:2002pp}
\bibinfo{author}{\bibfnamefont{C.}~\bibnamefont{Giunti}} \bibnamefont{and}
  \bibinfo{author}{\bibfnamefont{M.}~\bibnamefont{Tanimoto}},
  \bibinfo{journal}{Phys. Rev.} \textbf{\bibinfo{volume}{D66}},
  \bibinfo{pages}{113006} (\bibinfo{year}{2002}), \eprint{hep-ph/0209169}.

\bibitem{Antusch:2005kw}
\bibinfo{author}{\bibfnamefont{S.}~\bibnamefont{Antusch}} \bibnamefont{and}
  \bibinfo{author}{\bibfnamefont{S.~F.} \bibnamefont{King}},
  \bibinfo{journal}{Phys. Lett.} \textbf{\bibinfo{volume}{B631}},
  \bibinfo{pages}{42} (\bibinfo{year}{2005}), \eprint{hep-ph/0508044}.

\bibitem{Kang:2005as}
\bibinfo{author}{\bibfnamefont{S.~K.} \bibnamefont{Kang}},
  \bibinfo{author}{\bibfnamefont{C.~S.} \bibnamefont{Kim}}, \bibnamefont{and}
  \bibinfo{author}{\bibfnamefont{J.}~\bibnamefont{Lee}},
  \bibinfo{journal}{Phys. Lett.} \textbf{\bibinfo{volume}{B619}},
  \bibinfo{pages}{129} (\bibinfo{year}{2005}), \eprint{hep-ph/0501029}.

\bibitem{Cheung:2005gq}
\bibinfo{author}{\bibfnamefont{K.}~\bibnamefont{Cheung}},
  \bibinfo{author}{\bibfnamefont{S.~K.} \bibnamefont{Kang}},
  \bibinfo{author}{\bibfnamefont{C.~S.} \bibnamefont{Kim}}, \bibnamefont{and}
  \bibinfo{author}{\bibfnamefont{J.}~\bibnamefont{Lee}},
  \bibinfo{journal}{Phys. Rev.} \textbf{\bibinfo{volume}{D72}},
  \bibinfo{pages}{036003} (\bibinfo{year}{2005}), \eprint{hep-ph/0503122}.

\bibitem{Chauhan:2006im}
\bibinfo{author}{\bibfnamefont{B.~C.} \bibnamefont{Chauhan}},
  \bibinfo{author}{\bibfnamefont{M.}~\bibnamefont{Picariello}},
  \bibinfo{author}{\bibfnamefont{J.}~\bibnamefont{Pulido}}, \bibnamefont{and}
  \bibinfo{author}{\bibfnamefont{E.}~\bibnamefont{Torrente-Lujan}},
  \bibinfo{journal}{Eur. Phys. J.} \textbf{\bibinfo{volume}{C50}},
  \bibinfo{pages}{573} (\bibinfo{year}{2007}), \eprint{hep-ph/0605032}.

\bibitem{Rodejohann:2003sc}
\bibinfo{author}{\bibfnamefont{W.}~\bibnamefont{Rodejohann}},
  \bibinfo{journal}{Phys. Rev.} \textbf{\bibinfo{volume}{D69}},
  \bibinfo{pages}{033005} (\bibinfo{year}{2004}), \eprint{hep-ph/0309249}.

\bibitem{Li:2005ir}
\bibinfo{author}{\bibfnamefont{N.}~\bibnamefont{Li}} \bibnamefont{and}
  \bibinfo{author}{\bibfnamefont{B.-Q.} \bibnamefont{Ma}},
  \bibinfo{journal}{Phys. Rev.} \textbf{\bibinfo{volume}{D71}},
  \bibinfo{pages}{097301} (\bibinfo{year}{2005}), \eprint{hep-ph/0501226}.

\bibitem{Xing:2005ur}
\bibinfo{author}{\bibfnamefont{Z.-z.} \bibnamefont{Xing}},
  \bibinfo{journal}{Phys. Lett.} \textbf{\bibinfo{volume}{B618}},
  \bibinfo{pages}{141} (\bibinfo{year}{2005}), \eprint{hep-ph/0503200}.

\bibitem{Datta:2005ci}
\bibinfo{author}{\bibfnamefont{A.}~\bibnamefont{Datta}},
  \bibinfo{author}{\bibfnamefont{L.}~\bibnamefont{Everett}}, \bibnamefont{and}
  \bibinfo{author}{\bibfnamefont{P.}~\bibnamefont{Ramond}},
  \bibinfo{journal}{Phys. Lett.} \textbf{\bibinfo{volume}{B620}},
  \bibinfo{pages}{42} (\bibinfo{year}{2005}), \eprint{hep-ph/0503222}.

\bibitem{Everett:2005ku}
\bibinfo{author}{\bibfnamefont{L.~L.} \bibnamefont{Everett}},
  \bibinfo{journal}{Phys. Rev.} \textbf{\bibinfo{volume}{D73}},
  \bibinfo{pages}{013011} (\bibinfo{year}{2006}), \eprint{hep-ph/0510256}.

\bibitem{Winter:2007yi}
\bibinfo{author}{\bibfnamefont{W.}~\bibnamefont{Winter}},
  \bibinfo{journal}{Phys. Lett.} \textbf{\bibinfo{volume}{B659}},
  \bibinfo{pages}{275} (\bibinfo{year}{2008}), \eprint{arXiv:0709.2163
  [hep-ph]}.

\bibitem{Plentinger:2008up}
\bibinfo{author}{\bibfnamefont{F.}~\bibnamefont{Plentinger}},
  \bibinfo{author}{\bibfnamefont{G.}~\bibnamefont{Seidl}}, \bibnamefont{and}
  \bibinfo{author}{\bibfnamefont{W.}~\bibnamefont{Winter}},
  \bibinfo{journal}{JHEP} \textbf{\bibinfo{volume}{04}}, \bibinfo{pages}{077}
  (\bibinfo{year}{2008}), \eprint{0802.1718}.

\bibitem{Plentinger:2008nv}
\bibinfo{author}{\bibfnamefont{F.}~\bibnamefont{Plentinger}} \bibnamefont{and}
  \bibinfo{author}{\bibfnamefont{G.}~\bibnamefont{Seidl}}
  (\bibinfo{year}{2008}), \eprint{arXiv:0803.2889 [hep-ph]}.

\bibitem{Lee:1974jb}
\bibinfo{author}{\bibfnamefont{T.~D.} \bibnamefont{Lee}},
  \bibinfo{journal}{Phys. Rept.} \textbf{\bibinfo{volume}{9}},
  \bibinfo{pages}{143} (\bibinfo{year}{1974}).

\bibitem{Robinson:1997ub}
\bibinfo{author}{\bibfnamefont{M.~M.} \bibnamefont{Robinson}} \bibnamefont{and}
  \bibinfo{author}{\bibfnamefont{J.}~\bibnamefont{Ziabicki}}
  (\bibinfo{year}{1997}), \eprint{hep-ph/9705418}.

\bibitem{Kanemura:2007yy}
\bibinfo{author}{\bibfnamefont{S.}~\bibnamefont{Kanemura}} \emph{et~al.},
  \bibinfo{journal}{Eur. Phys. J.} \textbf{\bibinfo{volume}{C51}},
  \bibinfo{pages}{927} (\bibinfo{year}{2007}), \eprint{arXiv:0704.0697
  [hep-ph]}.

\bibitem{King:2005bj}
\bibinfo{author}{\bibfnamefont{S.~F.} \bibnamefont{King}},
  \bibinfo{journal}{JHEP} \textbf{\bibinfo{volume}{08}}, \bibinfo{pages}{105}
  (\bibinfo{year}{2005}), \eprint{hep-ph/0506297}.

\bibitem{Jenkins:2007ip}
\bibinfo{author}{\bibfnamefont{E.}~\bibnamefont{Jenkins}} \bibnamefont{and}
  \bibinfo{author}{\bibfnamefont{A.~V.} \bibnamefont{Manohar}},
  \bibinfo{journal}{Nucl. Phys.} \textbf{\bibinfo{volume}{B792}},
  \bibinfo{pages}{187} (\bibinfo{year}{2008}), \eprint{arXiv:0706.4313
  [hep-ph]}.

\bibitem{Jarlskog:1985ht}
\bibinfo{author}{\bibfnamefont{C.}~\bibnamefont{Jarlskog}},
  \bibinfo{journal}{Phys. Rev. Lett.} \textbf{\bibinfo{volume}{55}},
  \bibinfo{pages}{1039} (\bibinfo{year}{1985}).

\bibitem{Schwetz:2006dh}
\bibinfo{author}{\bibfnamefont{T.}~\bibnamefont{Schwetz}},
  \bibinfo{journal}{Phys. Scripta} \textbf{\bibinfo{volume}{T127}},
  \bibinfo{pages}{1} (\bibinfo{year}{2006}), \eprint{hep-ph/0606060}.

\bibitem{Huber:2006vr}
\bibinfo{author}{\bibfnamefont{P.}~\bibnamefont{Huber}},
  \bibinfo{author}{\bibfnamefont{J.}~\bibnamefont{Kopp}},
  \bibinfo{author}{\bibfnamefont{M.}~\bibnamefont{Lindner}},
  \bibinfo{author}{\bibfnamefont{M.}~\bibnamefont{Rolinec}}, \bibnamefont{and}
  \bibinfo{author}{\bibfnamefont{W.}~\bibnamefont{Winter}},
  \bibinfo{journal}{JHEP} \textbf{\bibinfo{volume}{05}}, \bibinfo{pages}{072}
  (\bibinfo{year}{2006}), \eprint{hep-ph/0601266}.

\bibitem{Ardellier:2006mn}
\bibinfo{author}{\bibfnamefont{F.}~\bibnamefont{Ardellier}} \emph{et~al.}
  (\bibinfo{collaboration}{Double Chooz})  (\bibinfo{year}{2006}),
  \eprint{hep-ex/0606025}.

\bibitem{Antusch:2004yx}
\bibinfo{author}{\bibfnamefont{S.}~\bibnamefont{Antusch}},
  \bibinfo{author}{\bibfnamefont{P.}~\bibnamefont{Huber}},
  \bibinfo{author}{\bibfnamefont{J.}~\bibnamefont{Kersten}},
  \bibinfo{author}{\bibfnamefont{T.}~\bibnamefont{Schwetz}}, \bibnamefont{and}
  \bibinfo{author}{\bibfnamefont{W.}~\bibnamefont{Winter}},
  \bibinfo{journal}{Phys. Rev.} \textbf{\bibinfo{volume}{D70}},
  \bibinfo{pages}{097302} (\bibinfo{year}{2004}), \eprint{hep-ph/0404268}.

\bibitem{Huber:2004gg}
\bibinfo{author}{\bibfnamefont{P.}~\bibnamefont{Huber}},
  \bibinfo{author}{\bibfnamefont{M.}~\bibnamefont{Lindner}}, \bibnamefont{and}
  \bibinfo{author}{\bibfnamefont{W.}~\bibnamefont{Winter}},
  \bibinfo{journal}{JHEP} \textbf{\bibinfo{volume}{05}}, \bibinfo{pages}{020}
  (\bibinfo{year}{2005}), \eprint{hep-ph/0412199}.

\bibitem{Huber:2006wb}
\bibinfo{author}{\bibfnamefont{P.}~\bibnamefont{Huber}},
  \bibinfo{author}{\bibfnamefont{M.}~\bibnamefont{Lindner}},
  \bibinfo{author}{\bibfnamefont{M.}~\bibnamefont{Rolinec}}, \bibnamefont{and}
  \bibinfo{author}{\bibfnamefont{W.}~\bibnamefont{Winter}},
  \bibinfo{journal}{Phys. Rev.} \textbf{\bibinfo{volume}{D74}},
  \bibinfo{pages}{073003} (\bibinfo{year}{2006}), \eprint{hep-ph/0606119}.

\end{thebibliography}
\end{document}